# *Kernel Based Estimation of Spectral Risk Measures*

S. Biswas* and R. Sen

Indian Statistical Institute, Bangalore, Karnataka, 560059, India

Spectral risk measures (SRMs) belong to the family of coherent risk measures. A natural estimator for the class of SRMs has the form of L-statistics. Various authors have studied and derived the asymptotic properties of the empirical estimator of SRM. We propose a kernel based estimator of SRM. We investigate the large sample properties of general L-statistics based on independent and identically distributed and dependent observations and apply them to our estimator. We prove that it is strongly consistent and asymptotically normal. We compare the finite sample performance of our proposed kernel estimator with that of several existing estimators for different SRMs using Monte Carlo simulation. We observe that our proposed kernel estimator outperforms all the estimators. Based on our simulation study we have estimated the exponential SRM of four future indices namely Nikkei 225, Dax, FTSE 100 and Hang Seng. We also discuss the use of SRM in setting initial margin requirements of clearinghouses. Finally we perform a backtesting exercise of SRM.

## 1. Introduction

In the financial market, a risk measure is used to determine the amount of capital to be kept in reserve. The reason for this hold is to limit the risks taken by financial establishments, such as banks and insurance agencies, so that it is acceptable to the regulator. A risk measure is a function that assigns real numbers to the possible outcomes of a random financial quantity, such as an insurance claim or loss of a portfolio. In recent years attention has turned towards convex and coherent risk measures, which have become increasingly popular in finance, insurance, and other areas associated with optimization under uncertainty. The concept of coherent risk measure was introduced by Artzner [1997], Artzner et al. [1999](see Appendix). Spectral risk measure (SRM) proposed by Acerbi [2002] is a weighted average of the quantiles of a loss distribution, the weights of which depend on the user's risk aversion. It belongs to the family of coherent risk measures. One nice feature of SRMs is that they relate the risk measure to the user's risk aversion (see Dowd et al. [2008]). In other words, if two investors are faced with the same distribution of potential misfortunes, an SRM shows that the more risk-averse investor faces a higher risk. Also, SRMs satisfy two additional conditions, i.e. law invariance and comonotonicity (see Henryk and Silvia [2008]). Law invariance is an important property for applications as it is a necessary property for a risk measure to be estimable from empirical data (see Henryk and Silvia [2008]).

The SRMs are more appealing than value at risk (VaR) and expected shortfall (ES) because of the unique characteristic that it directly links to the user's risk aversion. Here, we examine SRMs based on an exponential risk aversion function, which describes the user's risk aversion in terms of a single

*Corresponding author. Email: suparnabiswas_vs@isibang.ac.in

parameter, the coefficient of absolute risk aversion. For comprehensive discussion on the selection of utility function underpinning these risk measures are provided by Dowd et al. [2008], Wächter and Mazzoni [2013], and Brandtner and Kürsten [2015]. This kind of SRM is therefore dependent on a single parameter, whose value should be easy to detect in theory. Contrarily, the VaR and the ES depend on the confidence level, a parameter whose 'best' value is difficult to ascertain. Again, Overbeck [2004] discussed how SRM can be used for capital allocation, Baule [2022] addressed the problem of minimizing the risk of an exposure to a small number of defaultable counter parties based on SRMs, in particular ES, and Cotter and Dowd [2006] suggested that SRMs could be used by futures clearinghouses to set margin requirements that reflect their corporate risk aversion.

Definition 1 Let, $\phi \in \mathcal{L}^1([0,1])$ be an admissible risk spectrum (see Appendix) and $X$ denote the random loss variable. Then the spectral risk measure, according to Henryk and Silvia [2008], is defined by

$$M_\phi = \int_0^1 \phi(u) F_X^{-1}(u) du, \tag{1}$$

where $\phi$ is called the Risk Aversion Function and $F_X^{-1}$ is the quantile function of $X$(see Appendix).

The Risk Aversion Function assigns different weights to different $p$-confidence levels of the right tail of the distribution of $X$. Henryk and Silvia [2008] mentioned that any rational investor can express their subjective risk aversion by drawing a different profile for the weight function $\phi$. It can be seen that if $\phi(u) = \frac{1}{1-p} 1_{p \leq u \leq 1}$ then $M_\phi$ is the ES which is a SRM. But VaR is not a SRM as it is not a coherent risk measure.

We are interested in the estimation of SRMs. A related quantity is the distortion risk measure (DRM). The estimation literature is richer for DRMs than for SRMs.

Definition 2 Tsukahara [2009] Denoting the distribution of losses by $F_{-X}$, a DRM is defined as

$$\begin{aligned} \rho_{D_\theta} &= \int_{[0,1]} F_X^{-1}(u) dD_\theta(u) \\ &= \int_{\mathbb{R}} x dD_\theta \circ F_X(x), \end{aligned} \tag{2}$$

for a given distortion function $D_\theta$ (see Appendix) and $\theta$.

Comparing (1) and (2) we get,

$$M_\phi(X) = \rho_{D_\theta}(X) \quad \text{iff} \quad D_\theta(u) = \int_0^u \phi(1-s) ds \quad \forall u. \tag{3}$$

For $\rho_{D_\theta}$ to be coherent, $D$ must be convex.

A DRM of the form (2) suggests a natural estimator which can be written in the form of an $L$-statistic. Suppose we have independent observations $X_1, \dots, X_n$ and let $X_{n1} \leq \cdots \leq X_{nn}$ be the order statistics. If we replace $F$ by the empirical distribution function (df) $\hat{F}_n$ in equation (2), then we get a linear function of the order values as an estimator of $\rho_D$ which we denote as $\hat{\rho}$

$$\hat{\rho} = \sum_{i=1}^n c_{ni} X_{ni},$$

where $c_{ni} = D(i/n) - D((i-1)/n)$.

Similarly, SRM of the form (1) suggests a natural estimator in the form of an $L$-statistic. If we replace $F$ by the empirical df $\hat{F}_n$ then we can write an estimator of SRM as $\widehat{M_\phi}$ (see Adam et al. [2008])

$$\widehat{M_\phi} = -\sum_{i=1}^{n} c_{ni} X_{ni}$$

where $c_{ni} = \int_{(i-1)/n}^{i/n} \phi(1-s)ds \geq 0$.

Various authors have studied and derived the asymptotic properties of $\hat{\rho}$. All these results regarding $\hat{\rho}$ can be easily converted into those regarding $\widehat{M_\phi}$ by using equation (3). Shorack [1972] derived the asymptotic properties of $\hat{\rho}$ for i.i.d. case. Wellner [1977b] established certain almost sure "nearly linear" bounds of $\hat{F}_n$ and its left continuous inverse. Wellner [1977a] established a strong version of the Glivenko-Cantelli theorem for the uniform empirical df and used it to establish the asymptotic property of $\hat{\rho}$ for i.i.d. case. Sen [1978] also established the asymptotic properties of $\hat{\rho}$ for i.i.d. case. Zwet [1980] generalized the results of Wellner [1977a] and Sen [1978] considering i.i.d case. According to VanZwet all smoothness conditions on $g$ and $J$ are unnecessary and the pointwise convergence of $J_n$ can be relaxed (for definition of $g$, $J$ and $J_n$ see section 3.2). Tsukahara [2014] established the asymptotic property of $\hat{\rho}$ considering stationary process.

From the previous studies we observe that there are no results for the estimators of $M_\phi$ that involve estimators of the df other than the empirical df. Swanepoel and Van Graan [2005] introduced a kernel df estimator based on non-parametric transformation of the data. Dutta and Biswas [2017] used the Swanepoel and Graan's kernel df estimator to estimate the VaR and observed that it outperforms the empirical estimator of VaR. In this paper our aim is to consider kernel df estimators in the estimation of $M_\phi$ and establish their asymptotic properties. We organize the paper in the following way. In section 2 we propose kernel based estimators of $M_\phi$ using usual kernel df and Swanepoel and Graan's df estimator. In section 3 and 4 we establish the asymptotic properties and in the Appendix we give the detailed proof of our results. In section 5 we compare the finite sample performance of both the kernel based estimators with that of empirical estimator using Monte Carlo simulation. The comparisons are repeated for different sample sizes, risk measures and four different models. In section 6 using our simulation study we estimate the exponential SRM of four future indices-that is Nikkei 225, Dax, FTSE 100 and Hang Seng based on the daily return data for the periods January 2, 2004 to December 31, 2008 and January 2, 2009 to January 2, 2019. For comparison purposes, we also present the results for an earlier period namely January 1, 1991 to December 31, 2003 since this is the period considered by Cotter and Dowd [2006]. In section 7 we perform backtesting of SRM. And finally in section 8 we discuss the findings.

## 2. Proposed Estimator

The kernel method introduced by Rosenblatt [1956] has received considerable attention in nonparametric estimation. Let, $X_1, \ldots, X_n$ be i.i.d. random variables. The kernel df estimator is defined as follows

$$F_{n,b}(x) = \frac{1}{nb} \sum_{i=1}^{n} \int_{-\infty}^{x} k\left(\frac{t - X_i}{b}\right) dt$$
$$= \frac{1}{n} \sum_{i=1}^{n} K\left(\frac{x - X_i}{b}\right).$$

Let, us consider the following assumptions:

*Assumption 1:* The kernel df $K$ is differentiable, with a bounded kernel density $k$ with zero mean and finite variance.
*Assumption 2:* $b$ is the smoothing parameter satisfying the condition $b \to 0$ and $nb \to \infty$ as $n \to \infty$.

Another kernel df estimator based on non-parametric transformation of the data was introduced by Swanepoel and Van Graan [2005]. They have showed that the asymptotic bias and mean squared error of the estimator are considerably smaller than that of $F_{n,b}$. Swanepoel and Graan's df estimator is defined as follows:

$$\widetilde{F}_n(x) = \frac{1}{n}\sum_{i=1}^{n} K\left(\frac{F_{n,b}(x) - F_{n,b}(X_i)}{\hat{b}}\right), \tag{4}$$

where $\hat{b} = cb^{\alpha},\ 1 \leq \alpha < 3$. Swanepoel and Van Graan [2005] suggested to use

$$\hat{b} = b = \left[\frac{375\sqrt{3}}{28\pi}\right]^{1/7} \sigma^{-4/7} n^{-1/7}, \tag{5}$$

where $c$ and $\alpha$ is equal to 1, $\sigma = \min\{S, IQR/1.349\}$, $S$ and $IQR$ are the sample standard deviation and inter quartile range respectively. Based on this we propose the following estimator for $M_{\phi}$,

$$\widetilde{M_{\phi}} = -\int_0^1 \phi(u)\widetilde{F}_n^{-1}(u)du. \tag{6}$$

Dutta and Biswas [2017] observe that the Swanepoel and Grann's df estimator provides substantially improved quantile estimates than several quantile estimators for $p$ (location of quantile) close to zero and sample size less than equal to 500. Based on the findings of Dutta and Biswas [2017] we expect our proposed estimator to perform better than $\widehat{M_{\phi}}$.

For comparison purpose, we define another estimator based on the kernel df estimator i.e. $F_{n,b}(x)$ as

$$\widehat{M_{\phi}^{b}} = -\int_0^1 \phi(u)F_{n,b}^{-1}(u)du. \tag{7}$$

This will be useful as an intermediate measure for establishing the consistency and asymptotic normality of $\widetilde{M_{\phi}}$.

## 3. Consistency

Our objective is to establish the consistency of $\widetilde{M_{\phi}}$. We shall do this in two steps. First we establish the consistency of $\widehat{M_{\phi}^{b}}$ and then we can use this to establish the consistency of $\widetilde{M_{\phi}}$. We establish the consistency of $\widehat{M_{\phi}^{b}}$ by following the techniques used by Shorack [1972] and Wellner [1977a]. Let, $\xi_1, \xi_2, \ldots$ be a sequence of independent and identically distributed uniform $(0,\ 1)$ random variables with df $F(F(t) = t)$ on $[0,\ 1]$. Although results are derived for uniform $(0, 1)$, they hold true for all cdfs by Remark (1). Let, $F_{n,b}$ denote the kernel df estimator defined as follows.

$$F_{n,b}(t) = \frac{1}{nb}\sum_{i=1}^{n}\int_{-\infty}^{t} k\left(\frac{x - \xi_i}{b}\right)dx,\ 0 \leq t \leq 1$$

where $k$ is the kernel density function and $b$ is the smoothing parameter satisfying Assumptions 1-2. We know that SRM can be written in the form of $L$-statistics. So, equation (7) can be written in the form of $L$-statistics. Let, $\mathcal{G}$ denote the set of left continuous functions on $(0,\ 1)$ that are of bounded variation on $(\delta,\ 1-\delta)$, for all $\delta \in (0, 1/2)$; fix $g \in \mathcal{G}$. Let, $c_{n1}, \ldots,\ c_{nn}$ for $n \geq 1$, be known constants. Then we consider a general $L$-statistics of the form

$$T_n = \frac{1}{n}\sum_{i=1}^{n} g(\xi_{ni})c_{ni}$$

where $0 \leq \xi_{n1} \leq \ldots \leq \xi_{nn} \leq 1$ denote the order statistics of $n$ i.i.d uniform $(0,\ 1)$ random variables.

*Remark* 1 If $g = f(I^{-1}),\ f \in \mathcal{G}$ for some distribution function $I$, then $T_n$ has the same distribution as $S_n = \frac{1}{n}\sum_{i=1}^{n} c_{ni} f(X_{ni})$, where $X_{n1} \leq \ldots \leq X_{nn}$ are the order statistics of a sample of size $n$ from $I$.

### 3.1. *Independent Observations*

In this section we establish a strong law for $T_n$ when the observations are independent. For $n \geq 1$, let us define functions $J_n$ on $[0,\ 1]$ by $J_n(t) = c_{ni}$ for $(i-1)/n < t \leq i/n$, where $1 \leq i \leq n$ and $J_n(0) = c_{n1}$. Next, for $0 \leq t \leq 1$, we define $\psi_n(t) = -\int_t^1 J_n dF$ so that $\frac{c_{ni}}{n} = \left[\psi_n\left(\frac{i}{n}\right) - \psi_n\left(\frac{(i-1)}{n}\right)\right]$. Then

$$\begin{aligned} T_n &= \int_0^1 g(F_{n,b}^{-1})J_n dF \qquad (8)\\ &= \sum_{i=1}^{n} g(\xi_{ni})\left[\psi_n\left(\frac{i}{n}\right) - \psi_n\left(\frac{(i-1)}{n}\right)\right]. \end{aligned}$$

We, set

$$\mu_n = \int_0^1 g J_n dF.$$

In order to prove our two important results, we define certain functions and assume certain properties.
For fixed $b_1\ , b_2 > 0$ and $M > 0$ define a "scores bounding function" $B$ by

$$B(t) = Mt^{-b_1}(1-t)^{-b_2},\ 0 < t < 1.$$

For $\delta > 0$ define

$$\begin{aligned} D(t) &= Mt^{-1+b_1+\delta}(1-t)^{-1+b_2+\delta},\ 0 < t < 1,\\ h(t) &= [t(1-t)]^{1-\delta/2},\ 0 < t < 1. \end{aligned}$$

Now, let $g$ be a fixed function in $\mathcal{G}$. Let us denote $J$ to be a fixed measurable function on $(0,\ 1)$ and set

$$\mu = \int_0^1 J g dF. \qquad (9)$$

*Assumption(A):(Boundedness).* Let $|g| \leq D$, all $|g_n| \leq D$, $|J| \leq B$ and all $|J_n| \leq B$ on (0, 1) and suppose that $\int_0^1 Bhd|g| < \infty$.
*Assumption(B):(Smoothness).* Except on a set of $t$'s of $|g|$-measure 0 we have both $J$ is continuous at $t$ and $J_n \to J$ uniformly in some small neighbourhood of $t$ as $n \to \infty$.

THEOREM 3.1 *If Assumptions 1, 2, and (A) hold, then*

$$\lim_{n\to\infty}(T_n - \mu_n) = 0 \ w.p.1.$$

**Proof:** For proof see Appendix.

If $J$ and $g$ satisfy Asumption(A) then $|\mu| < \infty$. We state a Corollary which is similar to Corollary 2 of Wellner [1977a].

COROLLARY 3.2 *If* $\lim_{n\to\infty} \mu_n = \mu_\infty$ *exists (with* $|\mu_\infty| < \infty$*) and Assumptions 1, 2 and (A) hold, then*

$$\lim_{n\to\infty} T_n = \mu_\infty \ w.p.1.$$

**Proof:** For proof see Appendix.

THEOREM 3.3 *If Assumptions 1, 2, (A) and (B) hold, then*

$$\lim_{n\to\infty} T_n = \mu \ w.p.1$$

*where* $\mu$ *is finite.*

**Proof:** For proof see Appendix.

From Theorem 3.3 we can say that $\widehat{M_\phi^b}$ in (7) proves to possess strong consistency under the very general conditions stated above. Similarly we can prove that our estimator $\widetilde{M_\phi}$ in (6) possess strong consistency under the very general conditions stated above.

### 3.2. *Dependent Case*

THEOREM 3.4 *Let,* $\{\xi_n, n \geq 1\}$ *be an ergodic stationary sequence and if Assumptions 1, 2 and (A) holds, then*

$$\lim_{n\to\infty}(T_n - \mu_n) = 0 \ w.p.1.$$

Above mentioned theorem is stated and proved in Theorem 3.1 for the i.i.d. case. The proof remains true when $\{\xi_n, n \geq 1\}$ is an ergodic stationary sequence because all we need is the strong law of large numbers and the almost sure convergence of the df estimator (see Tsukahara [2014]). Therefore Theorem 3.3 is also true if $\{\xi_n, n \geq 1\}$ is an ergodic stationary sequence and hence we have the following result.

THEOREM 3.5 *Let, $\{\xi_n, n \geq 1\}$ be an ergodic stationary sequence and if Assumption 1, 2, (A) and (B) holds, then*

$$\lim_{n\to\infty} T_n = \mu \ w.p.1$$

*where $\mu$ is finite.*

From Theorem 3.5 we can say that $\widehat{M^b_\phi}$ in (7) proves to possess strong consistency under the very general conditions stated above. Similarly we can prove that our estimator $\widetilde{M_\phi}$ in (6) possess strong consistency under the very general conditions stated above.

## 4. Asymptotic Normality

In this section we establish the asymptotic normality of $\widetilde{M_\phi}$. The technique is similar to Shorack [1972] and Tsukahara [2014]. In order to establish the asymptotic normality of $\widetilde{M_\phi}$, first we need to establish the uniform central limit theorems for transformed kernel density estimators. We use the technique of Giné and Nickl [2008]. We introduce certain notations and define certain functions to prove our result.

### 4.1. *Independent Observations*

Let, $X_1, \ldots, X_n$ be i.i.d random variables with common law $\mathbb{P}$ on $\mathbb{R}$ and let $\mathbb{L}_n = \frac{1}{n}\sum_{i=1}^n \delta_{X_i}$ be the corresponding empirical measure. For an arbitrary (non-empty) set $M$, $\ell^\infty(M)$ will denote the Banach space of bounded real-valued functions $H$ on $M$ normed by $||H||_M := \sup_{m\in M}|H(m)|$. We denote $\mathcal{B}_s$ the Borel $\sigma$-algebra of a topological space $S$. For $h : \mathbb{R} \to \mathbb{R}$ a Borel-measurable function and $\mu$ a Borel measure on $\mathbb{R}$, we set $\mu_h := \int_{\mathbb{R}} h d\mu$ and $||h||_{p,\mu} := (\int_{\mathbb{R}} |h|^p d\mu)^{1/p}$, $1 \leq p \leq \infty$. We write $\mathcal{L}^p(\mathbb{R}, \mu)$ for the space of all Borel-measurable functions $h : \mathbb{R} \to \mathbb{R}$ that satisfy $||h||_{p,\mu} < \infty$. The symbol $\lambda$ denotes the Lebesgue measure on $\mathbb{R}$.

The symbol $C(\mathbb{R})$ denotes the Banach space of bounded real-valued continuous functions on $\mathbb{R}$ normed by the usual sup-norm $||.||_\infty$. Let, $\alpha = (\alpha_1, \ldots, \alpha_d)$ be a multi-index of nonnegative integers $\alpha_i$, set $|\alpha| = \sum_{i=1}^d \alpha_i$, and let

$$D^\alpha = \frac{\partial^{|\alpha|}}{(\partial_{x_1})^{\alpha_1}\cdots(\partial_{x_d})^{\alpha_d}}$$

denote the partial differential operator of order $\alpha$. For any nonnegative integer $s$, $C^s(\mathbb{R})$ denotes the Banach space of all bounded continuous real-valued functions that are $s$-times continuously differentiable on $\mathbb{R}$, equipped with the norm

$$||f||_{s,\infty} = \sum_{0\leq|\alpha|\leq s} ||D^\alpha f||_\infty.$$

The symbol $BV(\mathbb{R})$ will denote the space of measurable functions $\mathbb{R} \mapsto \mathbb{R}$ of bounded variation, equipped with the total variation norm

$$||f||_{TV} = \sup\Big\{\sum_{i=1}^n |f(x_i) - f(x_{i-1})| : n \in \mathbb{N}, -\infty < x_1 < \cdots < x_n < +\infty\Big\}.$$

Definition 3 A kernel $k : \mathbb{R} \to \mathbb{R}$ of real order $r > 0$ is a Lebesgue integrable function, symmetric around the origin, such that $\int_{\mathbb{R}} k(y)dy = 1$, $\int_{\mathbb{R}} y^j k(y)dy = 0$ for $j = 1, \ldots, \{r\}$, and $\int_R |y|^r |k(y)|dy < \infty$ where $[r]$ is the largest integer strictly smaller than $r$.

If the bandwidth sequence $\hat{b}_n$ satisfies that $\hat{b}_n = o(1)$ as $n \to \infty$, then the transformed kernel density estimator is given by

$$\mathbb{L}_n * k = \frac{1}{n\hat{b}_n} \sum_{i=1}^{n} k\left(\frac{F_{\hat{b}_n}(x) - F_{\hat{b}_n}(X_i)}{\hat{b}_n}\right) \tag{10}$$

where $F_{\hat{b}_n} = \frac{1}{n}\sum_{i=1}^{n}\int_{-\infty}^{t} k(\frac{x-X_i}{\hat{b}_n})dx$.
**Condition 1:** Let the random variables $X_1, \ldots, X_n$ are i.i.d. according to the law $\mathbb{P}$ on $\mathbb{R}$ and $d\mathbb{P}(x) = p_0(x)d\lambda(x)$. The variables $X_i$ are taken to be the coordinate projections of the infinite product probability space $(\mathbb{R}^{\mathbb{N}}, \mathcal{B}_{\mathbb{R}^{\mathbb{N}}}, \mathbb{P}^{\mathbb{N}})$.

Theorem 4.1 *Let Condition 1 hold and suppose that $p_0$ is a bounded function, in which case we set $t_1 = 0$ in what follows, or assume $p_0 \in C^{t_1}(\mathbb{R})$ for some real $t_1 > 0$. Let, $\mathcal{U}$ be a bounded subset of $BV(\mathbb{R})$. Let, $k$ be a kernel of order $r = 2t_1 + 1 - l$ for some $l$, $0 \leq l < 2t_1 + 1$. If $\hat{b}_n > 0$ is such that $\hat{b}_n^{2t_1+2-l} n^{1/2} \to 0$ as $n \to \infty$, then*

$$\sqrt{n}(\mathbb{L}_n * k - \mathbb{P}) \rightsquigarrow_{\ell^\infty(\mathcal{U})} \mathcal{G},$$

*where $\mathcal{G}$ is the $\mathbb{P}$-Brownian bridge indexed by $\mathcal{U}$. Convergence in law of random elements in $\ell^\infty(\mathcal{U})$ is defined as in Dudley [2014] (see page. 94) and is denoted by $\rightsquigarrow_{\ell^\infty(\mathcal{U})}$.*

**Proof:** For proof see Appendix.

*Remark* 2 (Order of the kernel and MISE-optimal rates) We note that the parameter $l \geq 0$ allows for some flexibility in the choice of the order of the kernel. In the most interesting case $\hat{b}_n \simeq n^{-1/(4t_1+3)}$ (Swanepoel and Van Graan [2005]), where the transformed kernel estimator simultaneously achieves optimal rates of convergence in squared error loss, we have $\hat{b}_n^{2t_1+2-l} n^{1/2} \to 0$ as $n \to \infty$ if $0 \leq l < 1/2$, so the kernel has to be of order $r > 2t_1 + 1/2$.

We state the immediate corollary for the cumulative df of the kernel estimator.

Corollary 4.2 *Let, Condition 1 hold and suppose that $p_0$ is a bounded function (case $t_1 = 0$) or assume that $p_0 \in C^{t_1}(\mathbb{R})$ for some real $t_1 > 0$. Let, $k$ be a kernel of order $r > 2t_1 + 1/2$. Choose $\hat{b}_n > 0$ of order $\hat{b}_n \simeq n^{-1/(4t_1+3)}$. Define the cumulative df $\widetilde{F}_n(t_1) = \int_{-\infty}^{t_1}(\mathbb{L}_n * k)(x)dx$ as well as $F(t_1) = \int_{-\infty}^{t_1} p_0(x)dx$. Then*

$$\sqrt{n}(\widetilde{F}_n - F) \xrightarrow{d} \mathbb{G},$$

*where $\mathbb{G}$ is the $\mathbb{P}$-Brownian bridge. That is $\{\mathbb{G}(t_1) : 0 \leq t_1 < 1\}$ is a Gaussian process with zero mean and covariance function $\sigma(s,\ t_1) = E\mathbb{G}(s)\mathbb{G}(t_1) = s \wedge t_1 - st_1$.*

Now using the above result we establish the asymptotic property of $\widetilde{M_\phi}$. If equation (8) is written using $\widetilde{F}_n$, then we can write $T_n$ as $W_n$

$$W_n = \int_0^1 g(\widetilde{F}_n^{-1}) J_n dF.$$

THEOREM 4.3 *Let Assumptions (A) and (B) hold. Let $k$ be a kernel of order $r > 2t_1 + 1/2$, for real $t_1 > 0$. Choose $\hat{b} > 0$ of order $\hat{b} \simeq n^{-1/(4t_1+1)}$. Then*

$$\sqrt{n}(W_n - \mu_n) \xrightarrow{d} N(0, \sigma^2),$$

*where*

$$\sigma^2 = \int_0^1 \int_0^1 (s \wedge t_1 - st_1) J(s) J(t_1) dg(s) dg(t_1) < \infty.$$

**Proof:** For proof see Appendix.

From Theorem 4.3 we have the following corollary.

COROLLARY 4.4 *If Assumptions (A) and (B) hold. Let $k$ be a kernel of order $r > 2t_1 + 1/2$, for real $t_1 > 0$. Choose $\hat{b} > 0$ of order $\hat{b} \simeq n^{-1/(4t_1+1)}$. Then*

$$\sqrt{n}(\widetilde{M}_\phi - M_\phi) \xrightarrow{d} N(0, \sigma^2),$$

*where*

$$\sigma^2 = \int_0^1 \int_0^1 (s \wedge t_1 - st_1) J(s) J(t_1) dg(s) dg(t_1) < \infty.$$

### 4.2. *Dependent Case*

Let, $\{X_n, n \geq 1\}$ be a strictly stationary $\phi$-mixing sequence of random variables. By definition, $\{X_n, n \geq 1\}$ is $\phi$-mixing if for each $k > 0$ and for each $n \geq 1$, $A \in \mathcal{M}_0^k$ and $B \in \mathcal{M}_{k+n}^\infty$,

$$|\mathbb{P}(A \cap B) - \mathbb{P}(A)\mathbb{P}(B)| \leq \phi(n)\mathbb{P}(A),$$

where $\phi$ is a nonnegative function of positive integers and $\mathcal{M}_a^b$ denotes $\sigma$-algebra generated by $X_a, X_{a+1}, \ldots, X_b$, for $a \leq b$ (see Degenhardt et al. [1996]).

Let, $x \in \mathbb{R}$ be fixed and $C[0,1]$ be the space of all continuous functions on $[0,1]$ endowed with the uniform metric and introduce some regularity conditions.

$(A_x)$ (i) $f$ is differentiable with bounded derivative $f'$ on the support of $F$. $f'$ is continuous in a neighborhood of $x$ and $f'(x) \neq 0$.

(ii) $\int_{-\infty}^{\infty} tk(t)dt = 0$ and $\int_{-\infty}^{\infty} t^2 k(t)dt < \infty$.

(Z) There exists an $x' \in \mathbb{R}$ such that $f$ and $k$ satisfy $A_{x'}$ (i.e. $Z = \bigcup_{y \in \mathbb{R}} A_y$).

Our next result gives necessary and sufficient conditions for the functional central limit theorem to hold for the transformed kernel df estimator under $\phi$-mixing process which is similar to the result proved by Degenhardt et al. [1996] (see Theorem 2.3).

THEOREM 4.5 *Let, $\sum_0^\infty n^2(\phi(n))^{1/2} < \infty$.*
*(a) Suppose that $f$ and $k$ satisfy (Z). Then we have*

$$\sqrt{n}(\widetilde{F}_n - F) \xrightarrow{\mathcal{D}} U, \tag{11}$$

*iff* $n^{1/4}\hat{b}_n \to 0$, *where* $U$ *is the Gaussian random process on [0, 1] with* $EU(t) = 0$ *and*

$$EU(s)U(t_1) = \sigma(s,t_1) = s \wedge t_1 - st_1 + \sum_{k=1}^{\infty} Eg_s(U_1)g_{t_1}(U_{k+1}) + \sum_{k=1}^{\infty} Eg_s(U_{k+1})g_{t_1}(U_1).$$

*where* $g_{t_1}(x) = 1_{[0,t_1]}(x) - t_1$. *Symbol* $\xrightarrow{\mathcal{D}}$ *denote weak convergence in* $C[0,\ 1]$.
*(b) Suppose that* $||f|| < \infty$. *Then* $(11)$ *holds iff*

$$\forall \varepsilon > 0 : \sqrt{n} \sup_x \int_{|t_1| > \varepsilon/\sqrt{n}} |F(x - t_1) - F(x)| k(t_1) dt_1 \to 0$$

*holds.*

**Proof:** For proof see Appendix.

Now using the above result we establish the asymptotic property of $\widetilde{M_\phi}$ under $\phi$-mixing process.

THEOREM 4.6 *Let,* $\sum_0^\infty n^2(\phi(n))^{1/2} < \infty$ *and Assumptions (A) and (B) hold. Suppose that* $f$ *and* $k$ *satisfy (Z) and* $n^{1/4}\hat{b}_n \to 0$. *Then*

$$\sqrt{n}(W_n - \mu_n) \xrightarrow{d} N(0, \sigma^2),$$

*where*

$$\sigma^2 = \int_0^1 \int_0^1 \sigma(s,t_1) J(s) J(t_1) dg(s) dg(t_1) < \infty.$$

**Proof:** For proof see Appendix.

From Theorem 4.6 we have the following corollary.

COROLLARY 4.7 *Let,* $\sum_0^\infty n^2(\phi(n))^{1/2} < \infty$ *and Assumptions (A) and (B) hold. Suppose that* $f$ *and* $k$ *satisfy (Z) and* $n^{1/4}\hat{b}_n \to 0$. *Then*

$$\sqrt{n}(\widetilde{M_\phi} - M_\phi) \xrightarrow{d} N(0, \sigma^2),$$

*where*

$$\sigma^2 = \int_0^1 \int_0^1 \sigma(s,t_1) J(s) J(t_1) dg(s) dg(t_1) < \infty.$$

## 5. Simulation

In order to compare the behaviour of different estimators in finite samples, we compute the mean squared error (MSE) of the estimators by simulating observations from several models. We consider three models.

- $(i)$ $\{X_i\}_{i=1,2,\ldots}$is an i.i.d. process, marginal distribution GPD with $\xi = 1/3$.
- $(ii)$ $\{X_i\}_{i=1,2,\ldots}$is an i.i.d. process, marginal distribution student's-t with 4 df.
- $(iii)$ $\{X_i\}_{i=1,2,\ldots}$is an i.i.d. process, marginal distribution $N(0,1)$.

The first two models are driven by Cont's (Cont [2001]) empirical observations regarding the extent of tail heaviness of the marginal asset return distributions.

To study the effect of dependence on the above mentioned estimators of SRM we consider the following GARCH(1,1) model

$$(iv)\ X_i = \sigma_i Z_i,$$
$$\sigma_i^2 = 0.061 X_{i-1}^2 + 0.932\sigma_{i-1}^2.$$

The model (iv) is the GARCH model fitted to the Nifty 50 daily loss data for the duration 1st January 2009 to 1st January 2019. The data is collected from national stock exchange (NSE) website (NSE [2019]). There are 2476 daily log return values (log returns are calculated considering the closing value of the index) in our data.

It is difficult to compute the exact value of the MSE of these estimators even if the the data generating process is completely specified. Therefore we use Monte-Carlo (MC) simulation to approximate the MSE of each of these estimators. The Monte-Carlo (MC) estimate of the MSE of any estimator $P_n$ of a parameter $\Theta$ is defined as $\frac{1}{B}\sum_{j=1}^{B}(P_{nj}-\Theta)^2$, where $B$ is the number of MC samples each of size $n$ drawn from a given process and $P_{nj}$ is the estimate based on the $j$th MC sample, $j = 1, \cdots, B$. In our work we use $B = 1000$. In the following sections we present the results for different risk aversion functions.

### 5.1. *Exponential Odds Risk Aversion*

The Risk Aversion Function defined by Cotter and Dowd [2006] is

$$\phi(u) = \frac{\beta e^{-\beta(1-u)}}{1-e^{-\beta}} \tag{12}$$

where $\beta \in (0, \infty)$ is the user's coefficient of absolute risk aversion. The coefficient of absolute risk aversion $\beta$ plays an important role in SRMs which is similar to the role played by the confidence level in the VaR and ES. Cotter and Dowd [2006] mentioned that the higher is $\beta$, the more we care about the higher losses relative to the others.

We consider the expression given in equation (1) and use the risk aversion function defined in (12) and compare the MSE of four estimators of SRM in Table 1. The estimators are

(i) The empirical estimator $\widehat{M_\phi}$

(ii) The kernel estimator $\widehat{M_\phi^b}$ estimated using the plug-in bandwidth proposed by Altman and Leger [1995] defined as

$$h_{AL} = \Big(\frac{1/4\widehat{V}}{\widehat{B}}\Big)^{1/3} n^{-1/3},$$

where

$$\widehat{V} = \varrho(k)\frac{1}{n(n-1)}\sum_{i=1}^{n}\sum_{j=1, i\neq j}^{n}\frac{1}{\alpha}k\Big(\frac{x_i - x_j}{\alpha}\Big),$$

and

$$\widehat{B} = 0.25\widehat{D}(F)(\mu_2(k))^2,$$

where $\varrho(k) = 2\int_{-\infty}^{+\infty} xk(x)K(x)dx$, $\mu_2(k) = \int_{-\infty}^{+\infty} x^2k(x)dx$ and

$$\widehat{D} = \frac{1}{n^3\alpha_b^4}\sum_{i=1}^{n}\sum_{j=1}^{n}\sum_{l=1}^{n} k_b'\Big(\frac{x_i - x_j}{\alpha_b}\Big)k_b'\Big(\frac{x_i - x_l}{\alpha_b}\Big).$$

$k_b'$ is the derivative of a kernel function $k_b$ (not necessarily equal to $k$). In practice $\alpha_b = \alpha$ and $k_b = k$. If we use the kernel function as Epanechnikov kernel, Altman and Leger [1995] proved that an optimal choice is made by taking $\alpha = n^{-0.3}\hat{\sigma}(x_i)$, where $\hat{\sigma}(x_i) = \min\left\{\hat{s},\ \frac{Q_3 - Q_1}{1.349}\right\}$, with $\hat{s}$ the sample standard deviation, and $Q_1$, $Q_3$ denote the first and third quartile respectively. We have estimated $h_{AL}$ using the ALbw function in kerdiest package in R software.

(iii) Our proposed kernel estimator $\widetilde{M}_\phi$ using the bandwidth defined in equation (5).

(iv) The kernel estimator $\widehat{M}_\phi^Q$ estimated using (1), where $Q_u$ is the kernel quantile estimator proposed by Sheather and Marron [1990]. The kernel quantile estimator is defined as follows

$$Q_u = \sum_{i=1}^{n}\left[\int_{\frac{i-1}{n}}^{\frac{i}{n}} k\left(\frac{t-u}{b}\right)dt\right]X_{(i)},$$

and use the bandwidth $h_{AL}$.

In Table 1, we present the ratio of the MSE of the estimators $\widehat{M}_\phi^b$, $\widetilde{M}_\phi$ and $\widehat{M}_\phi^Q$ with respect to the MSE of $\widehat{M}_\phi$ for $\beta = 1,\ 5,\ 10$ and 20 and for $n = 30,\ 100$ and 250 considering the four models. We observe that the estimator $\widehat{M}_\phi^b$ performs better than the empirical estimator $\widehat{M}_\phi$ only in case of model (iii), for small sample and for higher $\beta$ values. The estimator $\widetilde{M}_\phi$ outperforms the empirical estimator $\widehat{M}_\phi$ for all the models mentioned above. The highest gains being for low $\beta$ values and low sample size. Also, the empirical estimator $\widehat{M}_\phi$ outperforms the estimator $\widehat{M}_\phi^b$. We can conclude that our proposed estimator $\widetilde{M}_\phi$ outperforms the empirical estimator as well as the kernel estimator $\widehat{M}_\phi^b$. It is important to note that $\widetilde{M}_\phi$ outperforms the kernel estimator $\widehat{M}_\phi^b$ in all cases except Normal distribution with high $\beta$ and low sample size.

### 5.2. *Proportional Odds Distortion*

In order to get a broader idea we have used the DRM form given in equation (2) and compare the empirical and kernel estimator of DRM. Tsukahara [2009] suggested a one-parameter family of distortion that yields several classes of coherent risk measures:

- Proportional hazards (PH) distortion: $D_\theta^{PH}(u) = 1 - (1-u)^\theta$ where $\theta > 0$ and $D_\theta^{PH}$ is convex if $0 < \theta \leq 1$.
- Proportional odds (PO) distortion: $D_\theta^{PO}(u) = \theta u/[1-(1-\theta)u]$ where $\theta > 0$ and $D_\theta^{PO}$ is convex if $0 < \theta \leq 1$.
- Gaussian distortion (GA): $D_\theta^{GA}(u) = \Phi(\Phi^{-1}(u) + log\theta)$ where $\Phi$ is the standard normal df. Tsukahara have mentioned that a clear region of $\theta$ for which $\rho_{D_\theta}$ is finite is not found.

Tsukahara [2009] concluded that PO distortion is the most promising because of its financial interpretability and numerical stability. So we estimate DRM for this distortion function using the empirical estimator $\rho_{D_\theta}^{PO}$ and the kernel estimator $\widetilde{\rho}_{D_\theta}^{PO}$, where $F$ is replaced by $\widetilde{F}_n$. In Table 2 we present the ratio of the MSE's of these two estimators for $\theta = 0.005,\ 0.025,\ 0.05$ and 0.10 and for $n = 30,\ 100$ and 250 considering the four models. We observe that the estimator $\widetilde{\rho}_{D_\theta}^{PO}$ outperforms the empirical estimator $\rho_{D_\theta}^{PO}$ in case of all the models and all the parameter values and sample sizes mentioned above.

### 5.3. *Expected Shortfall*

The final measure of risk that we want to estimate is the popular measure ES defined as

$$ES_p = \frac{1}{1-p}\int_p^1 Q(u)du.$$

Our proposed estimator $\widehat{ES}_{p,h}$ based on Swanepoel and Van Graan [2005] df estimator is defined as

$$\widehat{ES}_{p,h} = \frac{1}{1-p}\int_p^1 \widetilde{F}_n^{-1}(u)du.$$

Another estimator of ES proposed in Chen [2008] is as follows. Let $K$ be a kernel function, which is a symmetric probability density function, $G(t) = \int_t^\infty K(u)du$ and $G_h(t) = G(t/h)$, where $h$ is a positive smoothing bandwidth. The kernel estimator of the survival function $S(x) = 1 - F(x)$ is

$$S_h(z) = \frac{1}{n}\sum_{t=1}^n G_h(z - X_t).$$

A kernel estimator of $Q_p$, denoted as $\hat{q}_{p,h}$, is the solution of $S_h(z) = p$. The kernel estimator of ES proposed by Chen [2008] is given as

$$Chen_{p,h} = \frac{1}{np}\sum_{t=1}^n X_t G_h(\hat{q}_{p,h} - X_t).$$

We use the bandwidth proposed by Chen and Tang [2005] which is defined as follows,

$$h_{opt} = \left\{\frac{2f^3(Q_p)b_k}{\sigma_k^4(f^{(1)}(Q_p))^2}\right\}^{1/3} n^{-1/3},$$

where $b_k = \int uw(u)H(u)du$, and $\sigma_k^2 = \int u^2 w(u)du$. $H(\cdot)$ is the df of the distribution with density $w$. $h$ involves unknown constants $Q_p$, $f$ and its derivative $f^{(1)}$ at $Q_p$. Chen and Tang [2005] suggested to approximate $Q_p$ in $h$ by the corresponding sample quantile. The authors suggested to approximate $f$ and $f^{(1)}$ by the density and the first derivative of the generalized Pareto distribution.

In Table 3 we present the ratio of the MSE of the estimator $Chen_{p,h}$ to that of our estimator $\widehat{ES}_{p,h}$ for $p = 0.95,\ 0.97$ and $0.99$ and $n = 30,\ 100$ and $250$ considering the four models. We observe that the ratio is quite high for all models, parameter values and sample sizes.

## 6. Data Analysis

Our data set consists of daily log returns considering the end-of-day prices for heavily traded index futures that is, the FTSE100, DAX, Hang Seng and Nikkei225 futures, between January 1, 1991 and December 31, 2003. To compare the findings with that of Cotter and Dowd [2006] we have considered the daily negative log returns. There are 3280 daily negative log return values. Also we have considered recent data set for the same instruments, between January 2, 2004 to January 2, 2019. 2008 is considered to be the year where there was a huge financial crisis. So, in order to estimate the market risk and to see the effect of financial crisis in the market we break the data set from January 2, 2004 to December 31, 2008 and January 2, 2009 to January 2, 2019. There are 1250 and 2582 daily negative log return values. These data are collected from the macrotrends website (see Macrotrends [2019]). The first data set is similar to the data set considered by Cotter and Dowd [2006]. The Financial Times Stock Exchange 100 Index, also called the FTSE 100 Index is a share index of the 100 companies listed on the London Stock Exchange with the highest market capitalisation. It is seen as a gauge of prosperity for businesses regulated by UK company law. The DAX is a blue chip stock market index consisting of the 30 major German companies trading on the Frankfurt Stock Exchange. Prices are taken from the Xetra trading venue. The Hang Seng Index is a freefloat-adjusted market-capitalization-weighted stock-market index in Hong Kong. It is used to record and monitor daily changes of the largest companies of the Hong Kong stock market and is the main indicator of the overall market performance in Hong Kong. These 50 constituent companies represent about 58% of the capitalisation of the Hong Kong Stock Exchange. The Nikkei 225, more commonly called the Nikkei, the Nikkei index, or the Nikkei Stock Average, is a stock market index for the Tokyo Stock Exchange (TSE). It has been calculated daily by the Nihon Keizai Shinbun (The Nikkei) newspaper since 1950. It is a price-weighted index, operating in the Japanese Yen (JP¥), and its components are reviewed once a year.

We apply the kernel based estimator and estimate the exponential SRM of the FTSE100, DAX, Hang Seng and Nikkei225 futures index, for three periods. We also estimate the confidence intervals. We run $10,000$ simulated samples. In each iteration a random sample of size 100 is selected with replacement from the data. Using each sample, the lower and upper limits of the confidence intervals are calculated. In Table 4-6 we report the exponential SRM and the corresponding 90% confidence intervals of the SRM estimates of the data set representing three different periods. The $\beta$ values considered are 1, 5, 10, 20, 100 and 200. The main motive of using the $\beta$ values (i.e 20, 100 and 200) is to compare the findings with that of the findings of Cotter and Dowd [2006].

From Table 4 we have observed that FTSE100 is the least risky index and Hang Seng is the most risky index for the period January 1, 1991 to December 31, 2003. Similar observation is also seen in Cotter and Dowd [2006]. From Table 5 we observe that FTSE100 is the least risky index and Nikkei225 is the most risky index for the period January 2, 2004 to December 31, 2008. From Table 6 we observe that FTSE100 is the least risky index and Hang Seng is the most risky index for the period January 2, 2009 to January 2, 2019. From Table 4-6 it is observed that risk has decreased from the first and second period to the third across all the securities and all values of risk aversion and the variation has also reduced. From Table 5 it is observed that risk is high in the second period compared to first and third period. We have also observed that if we estimate the 90% confidence intervals as described in Cotter and Dowd [2006] we obtain similar type of results.

If we want to set the initial margin requirements, then for first, second and third period FTSE 100 will have the lowest margin requirements. Hang Seng will have the highest margin requirements for the first and third period. But for second period Nikkei225 has the highest margin requirements. Also we observe that risk has decreased from the first and second period to the third across all the securities. Hence we can say that initial margin requirements will also decrease across all the securities. We even observe that our SRM estimates have much narrower confidence intervals across all the securities and all the time periods as compared to the SRM estimates observed in Cotter and Dowd [2006]. Hence we can say that our SRM estimator is much more precise than the SRM estimator observed in Cotter and Dowd [2006].

## 7. Backtest

Risk measures play an important role in the computation of regulatory capital. The regulator must guarantee that the institution's risk calculation technique is conservative and that the resulting capital reserves are adequate. The backtesting approach is one of the most important quantitative methods used by regulators to examine the risk measurement methodology's con-servativeness. Backtesting involves generating the forecast of a risk measure based on a sample and then comparing the future observed loss to this forecast to determine a failure rate. In this section we perform a coverage backtest for SRM which was proposed by Costanzino and Curran [2015]. Let, $\{t_i\}_{i=0}^N$ be a sequence of historical trading days and $\{L_i\}_{i=1}^N$ the corresponding realized trading losses. For each trading day $i = 1, \dots, N$, let $VaR_i^\alpha$ denote the VaR at level $\alpha$ and $X_{VaR}^{(i)}(\alpha) := K\Big(\frac{F_{n,b}(VaR_i^\alpha) - F_{n,b}(L_i)}{b}\Big) \in [0, 1]$ denote the VaR failure, where $F_{n,b}$ is the kernel df estimator. The $VaR_i^\alpha$ is estimated using the estimator proposed by Dutta and Biswas [2017] with data available up to day $i-1$. The VaR estimator proposed by Dutta and Biswas [2017] is defined as follows

$$VaR^\alpha = inf\{x : \widetilde{F}_n(x) \geq 1 - \alpha\}, \tag{13}$$

where $\widetilde{F}_n(x)$ is defined as in equation (4). We define the VaR failure rate $X_{VaR}^N(\alpha) \in [0, 1]$ for level $\alpha \in [0, 1]$ over $N$ trading days as

$$\begin{aligned} X_{VaR}^N(\alpha) &:= \frac{1}{N}\sum_{i=1}^{N} X_{VaR}^{(i)}(\alpha) \\ &= \frac{1}{N}\sum_{i=1}^{N} K\Big(\frac{F_{n,b}(VaR_i^\alpha) - F_{n,b}(L_i)}{b}\Big), \end{aligned}$$

where $K\Big(\frac{F_{n,b}(VaR_i^\alpha) - F_{n,b}(L_i)}{b}\Big) = \frac{1}{h}\int_{-\infty}^{VaR_i^\alpha} k\Big(\frac{F_{n,b}(x) - F_{n,b}(L_i)}{b}\Big)dx.$

Similarly, we define the SRM failure rate $X_{SR}^N$.

Definition 4 (SRM failure rate) For an admissible risk spectrum $\phi$, let $X_{SR}^{(i)}(\phi) \in [0, 1]$ be defined by

$$X_{SR}^{(i)}(\phi) = \int_0^1 \phi(p) K\Big(\frac{F_{n,b}(VaR_i^p) - F_{n,b}(L_i)}{b}\Big)dp,$$

where $\phi(p)$ is defined in equation (12). We define the SRM failure rate $X_{SR}^N(\phi) \in [0, 1]$ for admissible risk spectrum $\phi$ as

$$\begin{aligned} X_{SR}^N(\phi) &:= \frac{1}{N}\sum_{i=1}^{N} X_{SR}^{(i)}(\phi) \\ &= \frac{1}{N}\sum_{i=1}^{N}\int_0^1 \phi(p) K\Big(\frac{F_{n,b}(VaR_i^p) - F_{n,b}(L_i)}{b}\Big)dp. \end{aligned} \tag{14}$$

Definition 5 (Null Hypothesis for Coverage Test) The null-hypothesis for the SRM Coverage Test is

$H_0$ : SRM failure i.e. $X_{SR}^{(i)}(\phi)$ observed at two different dates must be independently distributed i.e $\forall\ i \neq j$, and $P(L_i \leq VaR_i^p) = p\ \forall\ p \in supp\phi$.

Proposition 7.1 *(Mean and Variance of $X_{SR}^N(\phi)$ under $H_0$) We consider the random variable*

*$X^N_{SR}(\phi) \in [0,\ 1]$ defined by (14). Then under the null-hypothesis,*

$$\mu_\phi := E[X^N_{SR}(\phi)] = \int_0^1 \phi(p)p dp \tag{15}$$

*and*

$$\begin{aligned}\sigma^2_\phi &:= V[X^N_{SR}(\phi)] \\ &= \frac{1}{N}\Big(2\int_0^1\int_0^p \phi(p)\phi(q)q dp dq - \Big(\int_0^1 \phi(p)p dp\Big)^2\Big)\end{aligned} \tag{16}$$

**Proof:** For proof see Appendix.

Lemma 7.2 *The SRM Failure Rate $X^N_{SR}$ is asymptotically normal under the null hypothesis and therefore admits a Z-test.*

**Proof:** For proof see Appendix.

Theorem 7.3 *(Coverage Test for SRMs) Let $\mu_\phi$ and $\sigma_\phi$ be the mean and standard deviation of the SRM Failure Rate $X^N_{SR}(\phi)$ (defined in (14)) under the null hypothesis given by*

$$\mu_\phi = \int_0^1 \phi(p)p dp$$

$$\sigma^2_\phi = \frac{1}{N}\Big(2\int_0^1\int_0^p \phi(p)\phi(q)q dp dq - \Big(\int_0^1 \phi(p)p dp\Big)^2\Big)$$

*Then the Z-score $Z^N_{SR}$ defined by*

$$Z^N_{SR}(\phi) := \frac{X^N_{SR}(\phi) - \mu_\phi}{\sigma_\phi} \tag{17}$$

*defines a Z-test for $X^N_{SR}$ and a coverage test for $M_\phi$.*

**Proof:** For proof see Appendix.

### 7.1. *Simulation*

The simulated data used for the evaluation of the backtest is same as that in section 5. We keep the estimation learning period fixed and equal to one year (250 observations). The backtesting period is equal to 1-day, and rolling window length is set to 250. Thus, for a single test run we need 500 consequent observations to perform the backtesting exercise. Starting from 251 for each day VaR is estimated. We do that up to day 500. Let, $x = (x_1, \dots, x_{500})$ be the 500 observations. For $i = 1, \dots, 250$, the $i$-th backtesting day VaR are equal to

$$VaR^\alpha_i(x) = VaR^\alpha_{250}(x_i, \dots, x_{i+249}),$$

where the VaR estimator is defined in equation (13).

In our setting the test statistic given in equation (17) is given by

$$Z_{SR}^{250}(\phi) := \frac{X_{SR}^{250}(\phi) - \mu_\phi}{\sigma_\phi},$$

where $X_{SR}^{250}(\phi) = \frac{1}{250}\sum_{i=1}^{250}\int_0^1 \phi(p)K\left(\frac{F_{n,b}(VaR_i^\alpha(x))-F_{n,b}(x_{i+250})}{b}\right)dp$ and $\sigma_\phi^2 = \frac{1}{250}\left(2\int_0^1\int_0^p \phi(p)\phi(q)qdpdq - \left(\int_0^1 \phi(p)pdp\right)^2\right)$.

We estimate $Z_{SR}^{250}(\phi)$ for different values of $\beta = 1,\ 5,\ 10,\ 20,\ 100$, using the above mentioned simulated data. We simulate a Monte Carlo sample of size 1000, where each sample corresponds to 250 observations. In Table 7 we have estimated the $Z$ values for different values of $\beta$ and for different simulated data. From the Table it is observed that if the simulated data is GPD, Student's-t and $N(0,1)$ then for all values of $\beta$ we are accepting the null hypothesis and if the simulated data is GARCH(1,1) model we reject for higher values of $\beta = 10,\ 20$ and 100. This rejection is possibly due to the part of the null hypothesis that assumes independence, since the data generating mechanism gives rise to dependent data.

## 8. Conclusion

In reality, the argument over the best quantitative risk measure has largely centered on the differences between VaR, a quantile, and ES, a tail expectation. In this paper we have discussed about the SRMs, which are coherent in nature and they takes into account the user's risk aversion. We also talk about how SRM and DRM are equivalent. SRMs are much more useful in real-world situations than the traditional risk measures like VaR and ES, where VaR is not coherent and ES do not take into account the user's risk aversion. So it becomes important how efficiently we can estimate SRM and validate the SRM estimates.

We have proposed two types of kernel based estimator of SRM i.e. $\widehat{M}_\phi^b$ which is based on usual kernel df and $\widetilde{M}_\phi$ is based on Swanepoel and Graan's df estimator. We have derived the asymptotic properties of both the kernel based estimators of SRM, which have the form of $L$-statistics. The asymptotic results are based on i.i.d. case and stationary process. The kernel based estimators are strongly consistent and asymptotically normally distributed. We have also derived certain almost sure "nearly linear" bounds of the usual kernel df and Swanepoel and Graan's df estimator which plays an important role in establishing the strong consistency of the kernel based estimator of SRM.

From the simulation study we observe that the choice of the bandwidth, the choice of the $\beta$ in SRM and the choice of $\theta$ in DRM plays an important role. We observe that our proposed estimator $\widetilde{M}_\phi$ outperforms the empirical estimator and both the kernel estimators $\widehat{M}_\phi^b$ and $\widehat{M}_\phi^Q$. But for higher $\beta$ values and for small sample $\widehat{M}_\phi^b$ outperforms all the estimators if the simulated data is $N(0,1)$. Also, we compare our proposed kernel estimator $\widetilde{\rho}_{D_\theta}^{PO}$ using PO distortion function with that of empirical estimator $\rho_{D_\theta}^{PO}$ and observe that $\widetilde{\rho}_{D_\theta}^{PO}$ outperforms $\rho_{D_\theta}^{PO}$ in case of all the models and all values of $\theta$ and sample sizes considered in our simulation study. ES is a special case of SRM and hence we compare two estimators of ES viz, our proposed estimator $\widehat{ES}_{p,h}$ and the Chen's estimator $Chen_{p,h}$. We compare the ratio of MSE of $Chen_{p,h}$ with that of $\widehat{ES}_{p,h}$ and it shows that the ratio is very high in all the cases. Hence we can say that our proposed estimator outperforms all the estimators considered in our simulation study.

Based on our simulation study we estimate the exponential SRM of four heavily traded index futures that is, the FTSE100, DAX, Hang Seng and Nikkei225 futures considering the period from January 1, 1991 to December 31, 2003, January 2, 2004 to December 31, 2008 and from January 2, 2009 to January 2, 2019. The SRM estimates suggest that the FTSE100 is the least risky index and Hang Seng is the most risky index during the period from January 1, 1991 to December 31, 2003. Similar observation can also be seen in Cotter and Dowd [2006], where the authors have estimated the extreme spectral risk measures using the Peaks-over-threshold approach. For the period January 2, 2004 to December 31, 2008 we see that FTSE100 is the least risky index and Nikkei 225 is the most risky index. And for the period January 2, 2009 to January 2, 2019 we find that FTSE100 is the least risky index and Nikkei225 is the most risky index. It is also seen that the risk has decreased from the first (January 1, 1991 to December 31, 2003) and second (January 2, 2004 to December 31, 2008) period to the third (January 2, 2009 to December 31, 2019) across all the securities and all values of risk aversion and the variation has also reduced. We have used SRM in setting the initial margin requirement of clearinghouses. We also observe that our proposed SRM estimator is much more precise than the SRM estimator defined by Cotter and Dowd [2006].

Also a backtesting exercise of SRM has been performed using a Z-test. We observe that if the simulated data is a GARCH(1,1) model then we reject the null hypothesis for higher values of $\beta = 10,\ 20,\ 100$ which means that SRM is not estimated correctly by the respective method for higher values of $\beta$. Also we observe that if the simulated data is GPD, Student's-t and $N(0,1)$ we do not reject the null hypothesis for all the $\beta$ values which we have considered in our backtesting exercise. And hence we can say that SRM is estimated correctly by the respective method.

## Acknowledgments

We are thankful to the esteemed reviewer for the suggestions and important references which lead to significant improvement of the article.

## Funding

The research is supported by Department of Science and Technology, India grant number (SR/WOS-A/PM-37/2019) and Indian Statistical Institute, Bangalore, Karnataka, 560059, India.

## Conflict of interest

On behalf of the authors, the corresponding author states that there is no conflict of interest.

Table 1. Ratio of MSEs of different estimators to that of $\widehat{M}_\phi$.

| $\beta$ | n | GPD | Student t | $N(0,1)$ | GARCH | GPD | Student t | $N(0,1)$ | GARCH | GPD | Student t | $N(0,1)$ | GARCH |
|---|---|---|---|---|---|---|---|---|---|---|---|---|---|
| | | $\widehat{M}^b_\phi$ | | | | $\widetilde{M}_\phi$ | | | | $\widehat{M}^Q_\phi$ | | | |
| 20 | 30 | 1.3964 | 3.9774 | 0.6710 | 5.0869 | 0.9802 | 0.9824 | 0.9877 | 0.9904 | 1.3825 | 2.3039 | 1.6577 | 1.0029 |
| | 100 | 1.7148 | 4.0123 | 0.8020 | 5.8940 | 0.9847 | 0.9879 | 0.9940 | 0.9950 | 1.3300 | 1.7464 | 1.3715 | 1.0032 |
| | 250 | 1.9132 | 4.9281 | 1.0899 | 7.1349 | 0.9968 | 0.9972 | 0.9980 | 0.9981 | 1.5533 | 2.2036 | 1.3593 | 1.0290 |
| 10 | 30 | 1.4964 | 2.8382 | 0.8090 | 2.8869 | 0.9531 | 0.9598 | 0.9724 | 0.9758 | 1.2763 | 2.7011 | 1.7262 | 0.9716 |
| | 100 | 1.5517 | 3.2343 | 0.8926 | 3.3415 | 0.9706 | 0.9780 | 0.9882 | 0.9894 | 1.3121 | 1.5722 | 1.2870 | 1.0050 |
| | 250 | 1.6545 | 3.9987 | 1.0309 | 3.8686 | 0.9927 | 0.9937 | 0.9957 | 0.9957 | 1.4879 | 1.5541 | 1.2816 | 1.0280 |
| 5 | 30 | 1.3664 | 2.1305 | 0.9354 | 2.4043 | 0.9137 | 0.9143 | 0.9380 | 0.9424 | 1.1838 | 2.7148 | 1.9194 | 1.0060 |
| | 100 | 1.4728 | 2.6721 | 0.9189 | 2.3295 | 0.9536 | 0.9595 | 0.9760 | 0.9779 | 1.2002 | 1.4350 | 1.2426 | 1.0049 |
| | 250 | 1.6743 | 3.0750 | 1.0280 | 2.5977 | 0.9864 | 0.9872 | 0.9909 | 0.9910 | 1.4109 | 1.5512 | 1.2266 | 1.0190 |
| 1 | 30 | 1.4094 | 1.1690 | 0.9800 | 1.1244 | 0.7811 | 0.5500 | 0.6157 | 0.6785 | 1.0786 | 3.0934 | 2.5437 | 0.9948 |
| | 100 | 1.2204 | 1.6930 | 0.9832 | 1.5013 | 0.8906 | 0.7881 | 0.8551 | 0.8775 | 1.0988 | 1.3197 | 1.1907 | 1.0056 |
| | 250 | 1.9030 | 2.0120 | 0.9963 | 1.7569 | 0.9653 | 0.9260 | 0.9421 | 0.9495 | 1.3053 | 1.0397 | 1.1362 | 1.0101 |

Table 2. Ratio of MSEs estimated using $\rho^{PO}_{D_\theta}$ and $\widetilde{\rho}^{PO}_{D_\theta}$.

| $\theta$ | n | GPD | Student t | $N(0,1)$ | GARCH(1,1) |
|---|---|---|---|---|---|
| 0.10 | 30 | 0.9830 | 0.9883 | 0.9908 | 0.9942 |
| | 100 | 0.9826 | 0.9880 | 0.9968 | 0.9970 |
| | 250 | 0.9988 | 0.9992 | 0.9996 | 0.9996 |
| 0.05 | 30 | 0.9874 | 0.9914 | 0.9928 | 0.9956 |
| | 100 | 0.9872 | 0.9912 | 0.9975 | 0.9978 |
| | 250 | 0.9992 | 0.9994 | 0.9997 | 0.9997 |
| 0.025 | 30 | 0.9905 | 0.9933 | 0.9940 | 0.9966 |
| | 100 | 0.9904 | 0.9932 | 0.9979 | 0.9983 |
| | 250 | 0.9994 | 0.9996 | 0.9997 | 0.9997 |
| 0.005 | 30 | 0.9948 | 0.9960 | 0.9956 | 0.9980 |
| | 100 | 0.9947 | 0.9959 | 0.9984 | 0.9990 |
| | 250 | 0.9997 | 0.9997 | 0.9998 | 0.9998 |

Table 3. Ratio of MSEs estimated using $Chen_{p,h}$ and $\widehat{ES}_{p,h}$.

| $p$ | n | GPD | Student t | $N(0,1)$ | GARCH |
|---|---|---|---|---|---|
| 0.95 | 30 | 8.5642 | 11.1450 | 66.7714 | 22.7412 |
| | 100 | 6.3968 | 12.3971 | 67.6071 | 25.6631 |
| | 250 | 1.1136 | 1.8916 | 15.2275 | 16.8182 |
| 0.97 | 30 | 14.4229 | 25.5667 | 89.4630 | 20.2961 |
| | 100 | 8.2924 | 15.1187 | 71.7257 | 23.4684 |
| | 250 | 2.3966 | 4.3716 | 26.2397 | 24.1047 |
| 0.99 | 30 | 3.6366 | 5.4426 | 30.0641 | 7.7423 |
| | 100 | 1.9805 | 3.6561 | 32.0538 | 10.5667 |
| | 250 | 3.1954 | 4.2802 | 39.1366 | 19.4700 |

Table 4. Estimates of exponential spectral risk measure of future index (1/01/1991 – 31/12/2003) and 90% confidence intervals.

| Future Index | $\beta = 1$ | $\beta = 5$ | $\beta = 10$ | $\beta = 20$ | $\beta = 100$ | $\beta = 200$ |
|---|---|---|---|---|---|---|
| Nikkei | 0.239 | 0.964 | 1.507 | 2.266 | 6.667 | 11.651 |
| 225 | [0.127 0.351] | [0.858 1.070] | [1.395 1.619] | [2.148 2.384] | [6.552 6.782] | [11.538 11.764] |
| DAX | 0.202 | 0.912 | 1.476 | 2.284 | 6.856 | 11.892 |
| | [0.095 0.309] | [0.805 1.019] | [1.362 1.590] | [2.177 2.391] | [6.752 6.960] | [11.783 12.001] |
| FTSE | 0.162 | 0.729 | 1.199 | 1.912 | 6.314 | 11.332 |
| 100 | [0.089 0.235] | [0.654 0.804] | [1.125 1.273] | [1.828 1.996] | [6.239 6.389] | [11.251 11.413] |
| Hang | 0.217 | 0.980 | 1.581 | 2.433 | 7.189 | 12.369 |
| Seng | [0.099 0.334] | [0.858 1.102] | [1.461 1.701] | [2.311 2.555] | [7.064 7.314] | [12.248 12.490] |

Notes: Estimates are in daily % return.

Table 5. Estimates of exponential spectral risk measure of future index (2/01/2004 – 31/12/2008) and 90% confidence intervals.

| Future Index | $\beta = 1$ | $\beta = 5$ | $\beta = 10$ | $\beta = 20$ | $\beta = 100$ | $\beta = 200$ |
|---|---|---|---|---|---|---|
| Nikkei | 0.270 | 1.204 | 1.967 | 2.901 | 7.821 | 12.999 |
| 225 | [0.149 0.391] | [1.075 1.333] | [1.844 2.089] | [2.775 3.027] | [7.707 7.935] | [12.875 13.123] |
| DAX | 0.190 | 0.852 | 1.401 | 2.601 | 7.623 | 12.452 |
| | [0.086 0.294] | [0.762 0.942] | [1.294 1.508] | [2.495 2.707] | [7.528 7.718] | [12.360 12.544] |
| FTSE | 0.183 | 0.802 | 1.343 | 2.167 | 6.947 | 12.113 |
| 100 | [0.098 0.268] | [0.714 0.889] | [1.249 1.437] | [2.079 2.254] | [6.862 7.032] | [12.013 12.213] |
| Hang | 0.232 | 0.999 | 1.677 | 2.602 | 7.435 | 12.598 |
| Seng | [0.098 0.366] | [0.873 1.125] | [1.548 1.806] | [2.477 2.727] | [7.309 7.561] | [12.477 12.719] |

Table 6. Estimates of exponential spectral risk measure of future index (2/01/2009 – 2/01/2019) and 90% confidence intervals.

| Future Index | $\beta = 1$ | $\beta = 5$ | $\beta = 10$ | $\beta = 20$ | $\beta = 100$ | $\beta = 200$ |
|---|---|---|---|---|---|---|
| Nikkei | 0.185 | 0.847 | 1.381 | 2.152 | 6.694 | 11.777 |
| 225 | [0.092 0.278] | [0.751 0.943] | [1.293 1.469] | [2.057 2.247] | [6.600 6.788] | [11.677 11.877] |
| DAX | 0.171 | 0.797 | 1.305 | 2.048 | 6.472 | 11.506 |
| | [0.088 0.254] | [0.704 0.890] | [1.214 1.396] | [1.963 2.133] | [6.386 6.558] | [11.422 11.590] |
| FTSE | 0.156 | 0.704 | 1.165 | 1.868 | 6.217 | 11.225 |
| 100 | [0.084 0.228] | [0.633 0.775] | [1.092 1.238] | [1.795 1.941] | [6.145 6.289] | [11.156 11.294] |
| Hang | 0.187 | 0.846 | 1.378 | 2.153 | 6.736 | 11.843 |
| Seng | [0.101 0.273] | [0.764 0.928] | [1.296 1.460] | [2.075 2.231] | [6.655 6.817] | [11.763 11.923] |

Table 7. Z values for different $\beta$s.

| $\beta$ | GPD | Student t | $N(0,1)$ | GARCH(1,1) |
|---|---|---|---|---|
| 1 | 0.1316 | −0.4947 | −0.5659 | −0.0043 |
| 5 | −0.5170 | −0.2148 | −0.5386 | 0.4049 |
| 10 | −0.6514 | 0.0628 | −0.5885 | −2.2630 |
| 20 | −0.7327 | 0.3266 | −1.1340 | −2.7233 |
| 100 | −0.6114 | 0.4804 | −0.1530 | −2.6695 |

## Appendix A: Some Definitions

Definition 6 (Delbaen [2002]) Let, $\psi$ denote the real valued random variables on a probability space $(\Omega,\ \mathcal{F},\ \mathbb{P})$. A coherent risk measure is a function $\rho : \psi \to \mathbb{R}$ satisfying the following properties:

(i) $X \geq 0 \Rightarrow \rho(X) \leq 0$.
(ii) $X \geq Y \Rightarrow \rho(X) \leq \rho(Y),\ X, Y \in \psi$.
(iii) $\rho(\lambda X) = \lambda\rho(X),\ \forall\lambda \geq 0,\ X \in \psi$.
(iv) $\rho(X + k) = \rho(X) - k,\ \forall k \in \mathbb{R},\ X \in \psi$.
(v) $\rho(X + Y) \leq \rho(X) + \rho(Y),\ \forall X,\ Y \in \psi$.

Definition 7 Let, $X \in \psi$ and $F$ be the df of $X$ then $Q_p = \inf\{x : F(x) \geq 1 - p\},\ 0 < p < 1$ is the quantile function.

Definition 8 (Henryk and Silvia [2008]) An element $\phi \in \mathcal{L}^1([0, 1])$ is called an admissible risk spectrum if

(i) $\phi \geq 0$
(ii) $\int_0^1 |\phi(t)|dt = 1$
(iii) $\phi$ is non-decreasing.

Definition 9 (Henryk and Silvia [2008]) A function $D : [0, 1] \to [0, 1]$ is a distortion function if

(i) D(0)=0 and D(1)=1.
(ii) D is non-decreasing function.

Definition 10 A sequence $\{\mu_n\}_{n=1}^{\infty}$ of finite signed Borel measures on $\mathbb{R}$ is an approximate convolution identity if it converges weakly to point mass $\delta_0$ at 0. If, in addition, for every $a > 0$, $\lim_n |\mu_n|(\mathbb{R} \setminus [-a, a]) = 0$, then we call the sequence $\{\mu_n\}_{n=1}^{\infty}$ a proper approximate convolution identity.

## Appendix B: Convergence of $F_{n,b}$ and $\widetilde{F}_n$ to $F$

The convergence of $\hat{F}_n - F$, with respect to $d_h$-metric is an important tool in the study of linear rank statistics (Pyke and Shorack [1968]) and linear combinations of order statistics (Shorack [1972]). Similarly, the convergence of $F_{n,b} - F$ with respect to $d_h$-metric is an important tool in our analysis. If $h$ is a nonnegative function approaching zero at the endpoints of the interval [0, 1], and $x, y$ are functions on [0, 1], the $d_h$-metric is defined by $d_h(x, y) = d(x/h, y/h) = \sup_{0<t<1} |x(t) - y(t)|/h(t)$, where $d$ denotes the usual supremum metric. Now using the above definition we can define

$$d_h(F_{n,b}, F) = d(F_{n,b}/h, F/h) = \sup_{0\leq t\leq 1} |F_{n,b}(t) - t|/h(t).$$

From Winter [1973], we have

$$d(F_{n,b}, F) = \sup_{0\leq t\leq 1} |F_{n,b}(t) - t| \to 0 \text{ as } n \to \infty\ w.\ p.\ 1. \tag{B1}$$

In Theorem B.1 below we establish that $\int_0^1 (1/h)dF < \infty$ is both necessary and sufficient for $d_h(F_{n,b}, F) \to 0$ with probability one as $n \to \infty$. Here $\int \cdot\ dF$ denotes integration with respect to Lebesgue measure. Our main motive to establish this type of result is to provide strong laws of large numbers for linear functions of order statistics.

Definition 11 Let, $\mathcal{H}(\nearrow)$ denote the set of all nonnegative, nondecreasing, continuous functions

$h$ on [0, 1] for which $\int_0^1 (1/h)dF < \infty$. Let, $\mathcal{H}$ denote the set of all functions $h$ such that $h(t) = h(1-t) = \overline{h}(t)$ for $0 \le t \le 1/2$ and some $\overline{h}$ in $\mathcal{H}(\nearrow)$.

Theorem B.1 *Let, Assumption 1 and 2 hold.*
*(i) If $h \in \mathcal{H}(\nearrow)$ then*

$$\lim_{n\to\infty} d_h(F_{n,b}, F) = 0 \ w.p.1. \tag{B2}$$

*(ii) If $h$ is increasing on [0, 1] and $\int_0^1 (1/h)dF = +\infty$ then*

$$\limsup_{n\to\infty} d_h(F_{n,b}, 0) = +\infty \ w.p.1. \tag{B3}$$

**Proof:** For proof see Appendix section C.

*Remark* 3 (i) of the above Theorem may be extended, using symmetry

$$\lim_{n\to\infty} d_h(F_{n,b}, F) = \lim_{n\to\infty} d_h(F_{n,b} - F, 0) = 0$$

w.p.1 for $h \in \mathcal{H}$. Also, (i) implies that $\lim_{n\to\infty} d_h(F_{n,b}, 0) = d_h(F, 0)$ w.p.1 for $h \in \mathcal{H}(\nearrow)$.

*Remark* 4 For $h \in \mathcal{H}(\nearrow)$, we define a process $Y_i$ on $[0,1]$ where $Y_i(t) = \frac{K((t-\xi_i)/b)}{h(t)}$ and write $||l|| = d(l, 0)$ for $l \in G'[0,1] \equiv G'$ where $G'[0,1]$ is the set of right continuous functions on $[0,1]$ with left limits. Then $(G', ||\cdot||)$ is an (inseparable) Banach space, and (i) of Theorem (B.1) is a strong law of large numbers for Banach space valued random elements: $E(Y_1) = \frac{F + b^2 f' \mu_2(k)/2 + o(b^2)}{h} = F/h$, where $\mu_2(k) = \int_{-1}^1 z^2 k(z)dz$, $||Y_1|| = d_h(K((t-\xi_1)/b), 0) = K(0)/h(\xi_1)$, and (i) of Theorem B.1 asserts that if $E||Y_1|| = K(0)\int_0^1 (1/h)dF < \infty$, then

$$\lim_{n\to\infty} ||1/n \sum_1^n Y_i - E(Y_1)|| = 0 \ w.p.1.$$

We state a Corollary which is similar to Corollary 1 of Wellner [1977a].

Corollary B.2 *If $h \in \mathcal{H}(\nearrow)$ then for all $\tau > 1$*

$$P(F_{n,b}(t) > \tau d_h(F, 0)h(t) \ for \ some \ 0 < t \le 1 \ i.o.) = 0.$$

**Proof:** For proof see Appendix section C.

Wellner [1977b] proved certain almost sure "nearly linear" bounds for the empirical df $\hat{F}_n(t) = \frac{1}{n}\sum_{i=1}^n 1_{[0,t]}(\xi_i)$ for $0 \le t \le 1$ and $\hat{F}_n^{-1}$, the left continuous inverse of $\hat{F}_n$. In Theorem B.3 below, we derive these bounds for $F_{n,b}$.

Theorem B.3 *Let, Assumption 1 and 2 hold. Let, $\tau_1, \tau_2 > 1$ be fixed. Then there exists $0 < \lambda = \lambda(\tau_1, \tau_2) < 1/2$ and a set $A \subset \Omega$ with $P(A) = 1$ having the following properties: for all $\omega \in A$ there is an $N \equiv N(\omega, \tau_1, \tau_2)$ for which $n \ge N$ implies*

(i) $1 - \left(\frac{1-t}{\lambda}\right)^{1/\tau_2} \le F_{n,b}(t) \le (t/\lambda)^{1/\tau_1}$ for $0 \le t \le 1$,
(ii) $\lambda t^{\tau_1} \le F_{n,b}(t)$ for all $t$ such that $0 < F_{n,b}(t)$,
(iii) $F_{n,b}(t) \le 1 - \lambda(1-t)^{\tau_2}$ for all $t$ such that $F_{n,b}(t) < 1$,
(iv) $\lambda t^{\tau_1} \le F_{n,b}^{-1}(t) \le 1 - \lambda(1-t)^{\tau_2}$ for $0 \le t \le 1$,

(v) $F_{n,b}^{-1}(t) \le (t/\lambda)^{1/\tau_1}$ for $t \ge \frac{1}{n}$, and
(vi) $1 - \left(\frac{1-t}{\lambda}\right)^{1/\tau_2} \le F_{n,b}^{-1}(t)$ for $t \le 1 - \frac{1}{n}$.

**Proof:** For proof see Appendix section C.

Similarly we can show that Theorem B.1 and B.3 are also true for the df $\widetilde{F}_n$. We state two theorems below using $\widetilde{F}_n$.

Theorem B.4 *Let, Assumption 1 and 2 hold.*
*(i) If* $h \in \mathcal{H}(\nearrow)$ *then*

$$\lim_{n\to\infty} d_h(\widetilde{F}_n, F) = 0 \ w.p.1. \tag{B4}$$

*(ii) If* $h$ *is increasing on [0, 1] and* $\int_0^1 (1/h)dF = +\infty$ *then*

$$\limsup_{n\to\infty} d_h(\widetilde{F}_n, 0) = +\infty \ w.p.1. \tag{B5}$$

Theorem B.5 *Let, Assumption 1 and 2 hold. Let,* $\tau_1, \tau_2 > 1$ *be fixed. Then there exists* $0 < \lambda = \lambda(\tau_1, \tau_2) < 1/2$ *and a set* $A \subset \Omega$ *with* $P(A) = 1$ *having the following properties: for all* $\omega \in A$ *there is an* $N \equiv N(\omega, \tau_1, \tau_2)$ *for which* $n \ge N$ *implies*

(i) $1 - \left(\frac{1-t}{\lambda}\right)^{1/\tau_2} \le \widetilde{F}_n(t) \le (t/\lambda)^{1/\tau_1}$ for $0 \le t \le 1$,
(ii) $\lambda t^{\tau_1} \le \widetilde{F}_n(t)$ for all $t$ such that $0 < \widetilde{F}_n(t)$,
(iii) $\widetilde{F}_n(t) \le 1 - \lambda(1-t)^{\tau_2}$ for all $t$ such that $\widetilde{F}_n(t) < 1$,
(iv) $\lambda t^{\tau_1} \le \widetilde{F}_n^{-1}(t) \le 1 - \lambda(1-t)^{\tau_2}$ for $0 \le t \le 1$,
(v) $\widetilde{F}_n^{-1}(t) \le (t/\lambda)^{1/\tau_1}$ for $t \ge \frac{1}{n}$, and
(vi) $1 - \left(\frac{1-t}{\lambda}\right)^{1/\tau_2} \le \widetilde{F}_n^{-1}(t)$ for $t \le 1 - \frac{1}{n}$.

Theorem B.1 and Theorem B.3 play an important role in establishing a strong law for $T_n$ in section 3. Similarly Theorem B.4 and B.5 plays an important role in establishing the consistency of $\widetilde{M_\phi}$.

## Appendix C: Proofs

**Proof of Theorem B.1.** First we begin with (ii). Suppose that $h$ is increasing on [0, 1] and $\int_0^1 (1/h)dF = +\infty$. Then,

$$\begin{aligned} d_F(F_{n,b}, 0) &= \sup_{0\le t\le 1} (F_{n,b}(t)/t) \\ &\ge F_{n,b}(\xi_{n1})/\xi_{n1} \\ &= \frac{1}{n\xi_{n1}} \sum_{i=1}^{n} K\left(\frac{\xi_{n1} - \xi_i}{b}\right). \end{aligned}$$

From (i) of Theorem 1 of Robbins and Siegmund [1972] we see that if $\xi_1, \xi_2, \ldots$ are independent and uniform on (0, 1) and if $V_n = \min(\xi_1, \ldots, \xi_n) = \xi_{n1}$ and $c_n = 1/r$ for fixed $r$, where $r$ is any arbitrary. Then, if $c_n/n \downarrow$ for all sufficiently large $n$ and $\sum_{n=1}^{\infty} \frac{c_n}{n}$ diverges then $P(nV_n \le c_n \ i.o.) = 1$.

Hence, we can write

$$\limsup_{n\to\infty} d_F(F_{n,b}, 0) = +\infty \ w.p.1.$$

Now, if $h \leq aF$, for some $a > 0$ and using equation (B2) we have

$$\begin{aligned}
\limsup_{n\to\infty} d_h(F_{n,b}, 0) &= \limsup_{n\to\infty} \left( \sup_{0\leq t\leq 1} \left( \frac{F_{n,b}(t)}{h(t)} \right) \right) \\
&\geq \limsup_{n\to\infty} \left( \sup_{0\leq t\leq 1} \left( \frac{F_{n,b}(t)}{at} \right) \right) \\
&= \frac{1}{a} \limsup_{n\to\infty} d_F(F_{n,b}, 0) \\
&= +\infty \ w.p.1.
\end{aligned}$$

If $h \leq aF$ for some $a > 0$ does not hold, then for every $a > 0$, $h(t) > at$, for some $t \in [0,\ 1]$. Hence, by monotonicity of $h$ this implies that $h \geq aF$, for some $a > 0$.
Now, let $R_i(t) = \frac{1}{b} \int_{-\infty}^{t} k\left(\frac{x-\xi_i}{b}\right) dx$ so that

$$F_n = \frac{1}{n} \sum_{i=1}^{n} R_i(t).$$

Let, $M > 0$ and define events $B_n$ and $D_n$ by

$$B_n = \{d_h(F_{n,b}, 0) > M\} = \left\{ d_h \left( \sum_{i=1}^{n} R_i, 0 \right) > nM \right\}$$

and

$$D_n = \{d_h(R_n, 0) > nM\}$$

Now, since $\sum_{i=1}^{n} R_i \geq R_n$, $d_h\left(\sum_{i=1}^{n} R_i, 0\right) \geq d_h(R_n, 0)$ and hence we can write $\{D_n\ i\ .\ o\ .\} \subset \{B_n\ i\ .\ o\ .\}$. But the events $D_n$ are independent and therefore by Borel-Cantelli Lemma, we have

$$P(D_n\ i\ .\ o\ .) = 0 \ or\ 1\ according\ as \sum_{n=1}^{\infty} P(D_n) < \infty \ or\ = \infty \tag{C1}$$

Now, we compute $P(D_n)$. Since the $R_i$'s are independent and identically distributed we may drop the subscript $n$; hence for $n$ sufficiently large.

$$\begin{aligned}
P(D_n) &= P(d_h(R, 0) > nM) \\
&= P\left( \frac{K(0)}{h(\xi)} > nM \right) \ where,\ K(0) = \int_{-\infty}^{0} k(u)du \\
&= P(h(\xi) < K(0)n^{-1}M^{-1}) \\
&= P(\xi < h^{-1}(K(0)n^{-1}M^{-1})) \\
&= h^{-1}(K(0)n^{-1}M^{-1}).
\end{aligned}$$

Hence, the series in equation (C1) is $\sum_{n=1}^{\infty} h^{-1}(K(0)n^{-1}M^{-1})$ and this converges or diverges, by monotonicity, with $\int_0^{\infty} h^{-1}(K(0)t^{-1}M^{-1})dt$ and after change of variables we have $M^{-1}K(0)\int_0^{\infty} s^{-2}h^{-1}(s)ds$.
Now, integration by parts together with $h \geq aF$ shows that the latter integral converges and diverges with $\int_0^1 (1/h)dF$.

$$\begin{aligned}
\int_0^{\infty} s^{-2}h^{-1}(s)ds &= \int_0^1 s^{-2}h^{-1}(s)ds + \int_1^{\infty} s^{-2}h^{-1}(s)ds \\
&\leq \int_0^1 s^{-2}h^{-1}(s)ds + \int_1^{\infty} s^{-2}\frac{s}{a}ds \\
&\leq \left[s^{-2}\int h^{-1}(s)ds - \int(-2)s^{-3}\left(\int h^{-1}(s)ds\right)ds\right]_0^1 + \frac{1}{a}\int_1^{\infty}\frac{1}{s}ds \\
&= \left[s^{-2}\int h^{-1}(s)ds + 2\int s^{-3}\left(\int h^{-1}(s)ds\right)ds\right]_0^1 + \frac{1}{a}\int_1^{\infty}\frac{1}{s}ds
\end{aligned}$$

Hence, $\int_0^1 (1/h)dF = +\infty$ implies, by the divergence half of (C1), that $P(D_n\ i.o.) = 1$ and therefore $P(B_n\ i.o.) = 1$, for all $M > 0$. Since, $M$ is arbitary.
Hence, (ii) is proved. Note that we have also proved that $\int_0^1 (1/h)dF < \infty$ implies $P(d_h(R_n, 0) > nM\ i.o.) = 0$ for all $M > 0$. We now prove (i) Suppose, $h \in \mathcal{H}(\nearrow)$. Let, $\epsilon > 0$ and choose $\theta$ so small that $\int_0^{\theta}(1/h)dF < \epsilon/2$. Then

$$d_h(F_{n,b}, F) \leq \sup_{0<t\leq\theta}\left(\frac{F_{n,b}(t)}{h(t)}\right) + \sup_{0<t\leq\theta}\left(\frac{t}{h(t)}\right) + \sup_{\theta\leq t\leq 1}\frac{|F_{n,b}(t) - t|}{h(\theta)} \tag{C2}$$

$$\begin{aligned}
\sup_{0<t\leq\theta}\left(\frac{F_{n,b}(t)}{h(t)}\right) &= \sup_{0<t\leq\theta}\left(\frac{\frac{1}{nb}\sum_{i=1}^{n}\int_{-\infty}^{t} k\left(\frac{x-\xi_i}{b}\right)dx}{h(t)}\right) \\
&\leq \frac{\frac{1}{nb}\sum_{i=1}^{n}\int_{-\infty}^{\theta} k\left(\frac{x-\xi_i}{b}\right)dx}{h(\xi_i)} \\
&= \frac{\frac{1}{n}\sum_{i=1}^{n} K\left(\frac{\theta-\xi_i}{b}\right)}{h(\xi_i)} \\
&\to \int_0^{\theta}(1/h)dF\ w\ .\ p\ .\ 1\ \textit{by the ordinary strong law of large numbers.}
\end{aligned}$$

$\because \frac{t}{h(t)} \leq \int_0^t (1/h)dF$ which implies $\sup_{0<t\leq\theta}(t/h(t)) \leq \int_0^{\theta}(1/h)dF$. Now from equation (B1) we can say that the third term in equation (C2) converges to zero w. p. 1.
$\therefore$ We can write

$$\limsup_{n\to\infty} d_h(F_{n,b}, F) < \epsilon/2 + \epsilon/2 = \epsilon\ w\ .\ p\ .\ 1\ \textit{for any}\ \epsilon > 0.$$

Hence, (i) is proved.

**Proof of Corollary B.2.** From equation (B2) we can write that $d_h(F_{n,b}, 0) \to d_h(F, 0)$ w.p.1 a $n \to \infty$. Hence, for any $\tau > 1$, we can write

$$P(d_h(F_{n,b}, 0) > \tau d_h(F, 0)\ i.o.) = 0.$$

**Proof of Theorem B.3.** Note that it suffices to prove only the upper bound of (1) and (5): by replacing $\xi$ by $1 - \xi_i$, by interchanging $\tau_1$ and $\tau_2$, and by use of symmetry about the identity function, the upper bound of (1) implies the remaining inequalities in (1) and (4): similarly (5) implies the remaining inequalities (2), (3) and (6). The proof of (5) is similar to the proof of (8) in Theorem 1 from Wellner [1977b].
To prove the upper bound of (1), let $\alpha = 1/\tau_1$ and $\tau > 1$. We define $F^*_{n,b} = F_{n,b} - F$ and

$$E_n = \left\{ \sup_{0<t\leq 1} \frac{|F_{n,b}(t)|}{\tau t^\alpha} \geq 1 \right\}.$$

From Corollary 1.1 we can write that $P(E_n\ i.o.) = 0$. Hence for $n \geq N(\omega, \alpha)$,

$$|F_{n,b}(t)| \leq \tau t^\alpha,\ 0 \leq t \leq 1$$

or,

$$F_{n,b}(t) \leq (1 + \tau)t^\alpha,\ 0 \leq t \leq 1.$$

This implies that for $n \geq N(\omega, \alpha)$ and all $\omega$ in a set with probability one

$$F_{n,b}(t) \leq (t/\lambda)^\alpha,\ 0 \leq t \leq 1$$

where $0 < \lambda \equiv 2^{-1}(1 + \tau)^{-\tau_1} < \frac{1}{2}$. Hence the upper bound of (1) is proved.

**Proof of Theorem 3.1.** The proof is similar to the proof of Theorem 3 of Wellner [1977a].

**Proof of Corollary 3.2.** The proof is similar to the proof of Corollary 2 of Wellner [1977a].

**Proof of Theorem 3.3.** If we show that $\lim_{n\to\infty} \mu_n = \mu$, then Corollary 2 with $\mu_\infty = \mu$ is in force and the proof is complete. But, by Assumption(A) we have $|J_n g| \leq M^2[F(1 - F)]^{-1+\delta}$ which is in $L^1(F)$. Again from Assumption(B) we have $J_n(t)g(t) \to J(t)g(t)$ for all $t \in (0,\ 1)$. Therefore, by the dominated convergence theorem, we can write

$$\mu_n = \int_0^1 J_n g dF \to \int_0^1 J g dF = \mu.$$

Hence,

$$\lim_{n\to\infty} T_n = \mu\ w.p.1.$$

**Some Lemmas used in the proof of Theorem 5**
The symbol $C_0(\mathbb{R})$ denotes the closed subspace of $C(\mathbb{R})$ consisting of bounded continuous real-valued functions $f$ that vanish at $\infty$. The dual space of $C_0(\mathbb{R})$ denoted by $C_0(\mathbb{R})'$ is normed by the usual norm $||.||'_{C_0}$. Then $M(\mathbb{R}) = C_0(\mathbb{R})'$ is the space of signed Borel measures of finite variation on $\mathbb{R}$ and $||\mu||'_{C_0} = ||\mu||$, where $||\mu|| := |\mu|(\mathbb{R})$ is the total variation norm of $\mu$, $|\mu|$ being the total variation measure of $\mu \in M(\mathbb{R})$. The convolution of two signed Borel measures $\mu$ and $\nu$ on $\mathbb{R}$ is defined by $\mu * \nu(E) = \mu \times \nu(T^{-1}(E))$ where $E \in \mathcal{B}_{\mathbb{R}}$ and where $T : \mathbb{R} \times \mathbb{R} \to \mathbb{R}$ is addition $T(x, y) = x + y$, (see page 337 Giné and Nickl [2008]). If $f \in \mathcal{L}^p(\mathbb{R}, \lambda)$ $(1 \le p \le \infty)$ and $g \in \mathcal{L}^q(\mathbb{R}, \lambda)$ $(1 \le q \le \infty)$ with $1/p+1/q = 1$, then $f * g(x)$ defines an element of $C(\mathbb{R})$ and Young's inequality gives

$$||f * g||_\infty \le ||f||_{p,\lambda}||g||_{q,\lambda}. \tag{C3}$$

Again, if $g \in \mathcal{L}^p(\mathbb{R}, \lambda)$ $(1 \le p \le \infty)$ and $\mu \in M(\mathbb{R})$, then the function $g * \mu$ is well defined $\lambda$-a.e. and satisfies

$$||g * \mu||_{p,\lambda} = ||g||_{p,\lambda}||\mu||. \tag{C4}$$

If $f : \mathbb{R} \to \mathbb{R}$ is a locally integrable function, it gives rise to a distribution $T_f$ acting on the space $\mathcal{D}(\mathbb{R})$ of all infinitely differential real-valued functions on $\mathbb{R}$ with compact support via integration. We define the partial distributional derivative $D^\alpha_w f$ of a locally integrable function $f$ by the relation

$$D^\alpha_w T_f(\varphi) = (-1)^{|\alpha|} \int_{\mathbb{R}} f(x)(D^\alpha \varphi)(x)dx,$$

where $\varphi \in \mathcal{D}(\mathbb{R})$. If $D^\alpha_w T_f(\cdot)$ (for every $\alpha$, with $|\alpha| = 1$) is also a regular distribution given by another locally integrable function $g$, then we say that $g$ is the weak derivative of $f$, and we write $g = D^\alpha_w f$. If $f$ is differentiable then $D^\alpha = D^\alpha_w f$ holds $\lambda$-a.e. Also, we set $D_w f = D^\alpha_w f$ in case $\alpha = d = 1$.

We define a space $\mathcal{BV}(\mathbb{R})$ which is defined as follows

$$\mathcal{BV}(\mathbb{R}) = \{f : \mathbb{R} \to \mathbb{R} \text{ locally integrable}, D^\alpha_w f \in M(\mathbb{R}) \text{ for every } \alpha \text{ with} |\alpha| = 1\},$$

which means $\mathcal{BV}(\mathbb{R})$ is the space of all locally integrable functions that have weak partial derivatives of order one which are finite signed measures. The space $\mathcal{BV}(\mathbb{R})$ can be equipped with seminorm

$$||f||_{BV(\mathbb{R})} = \max_{\alpha:|\alpha|=1} ||D^\alpha_w f||,$$

where $||D^\alpha_w f|| = ||D^\alpha_w f||'_{C_0}$ is the total variation of the measure $D^\alpha_w f$. If $d = 1$, then $f \in \mathcal{BV}(\mathbb{R})$ iff there exists $g \in BV(\mathbb{R})$ such that $f = g$ holds $\lambda$-a.e. Again, if we consider $f \in \mathcal{BV}(\mathbb{R})$ and $\widetilde{f}(x) = D_w f(-\infty, x]$, then for all $x$, $y$ in a set $\Omega_f$ such that $\lambda(\Omega^c_f) = 0$, $x < y$, we have

$$|f(y) - f(x)| = |\widetilde{f}(y) - \widetilde{f}(x)| = |D_w f(x, y]| \le |D_w f|(x, y] \tag{C5}$$

and $\Omega^c_f$ is countable if $f \in BV(\mathbb{R})$. The above mentioned fact has been used by Giné and Nickl [2008], where $\nu_f$ is written for the measure $D_w f$. Again, by right-continuity of $\widetilde{f}$ and the definitions, we have

$$\begin{aligned}||\nu_f|| &= ||f||_{BV} = ||\widetilde{f}||_{BV} = ||\widetilde{f}||_{TV} \\ &= \sup\left\{\sum_{i=1}^{n}|f(x_i) - f(x_{i-1})| : n \in \mathbb{N}, -\infty < x_1 < \cdots < x_n < +\infty,\ x_i \in \Omega_f\right\} \\ &\leq ||f||_{TV},\end{aligned} \tag{C6}$$

where $||f||_{TV}$ may be infinite (if $f$ is in $\mathcal{BV}(\mathbb{R})$ but not in $BV(\mathbb{R})$).
For example, if $k \in \mathcal{L}^1(\mathbb{R}, \lambda)$ satisfies $\int_{\mathbb{R}} k(y)dy = 1$, and if $d\mu_n(y) = \frac{1}{\hat{b}_n} k\left(\frac{F_{\hat{b}_n}(y)}{\hat{b}_n}\right) dy$, for $0 < \hat{b}_n \to 0$, then the sequence $\mu_n$ is a proper approximate identity. Lemma 4 of Giné and Nickl [2008] can be written as

Lemma C.1 *Let $d\mu_{\hat{b}}(x) = \frac{1}{\hat{b}} k\left(\frac{F_{\hat{b}}(x)}{\hat{b}}\right) d\lambda(x)$ where $k$ is a kernel of order of $m \geq 0$ and $d\mathbb{P}(x) = p_0(x)d(x)$ be a probability measure with a density $p_0 \in \mathcal{L}^q(\mathbb{R}, \lambda)$. Let, $\emptyset \neq \mathcal{F} \subseteq \mathcal{L}^2(\mathbb{R}, \mathbb{P})$. Let further $f \in \mathcal{L}^p(\mathbb{R}, \lambda)$ where $1/p + 1/q = 1$ and set $\bar{f}(x) = f(-x)$ for $x \in \mathbb{R}$. Then $\bar{f} * p_0 \in C(\mathbb{R})$ and we have for the bias term that*

$$|E(f * \mu_{\hat{b}}(x) - f(X))| = \left|\int_{\mathbb{R}} \frac{k(t_2)[\bar{f} * p_0(F_g^{-1}(\hat{b}t_2)) - \bar{f} * p_0(0)]dt_2}{F'_{\hat{b}}(F_{\hat{b}}^{-1}(\hat{b}t_2))}\right| \text{ where } F'_{\hat{b}}(F_{\hat{b}}^{-1}(\hat{b}t_2)) \neq 0.$$

**Proof.** From equation(C4), $f * k \in \mathcal{L}^p(\mathbb{R}, \lambda)$ implies that $(f * k) \cdot p_0$ is Lebesgue-integrable for every $f \in \mathcal{F}$ since $p_0 \in \mathcal{L}^q(\mathbb{R}, \lambda)$. Now, by change of variables and Fubini's theorem, we have

$$\begin{aligned}E(f * \mu_h(X) - f(X)) &= \frac{1}{h}\int_{\mathbb{R}}\int_{\mathbb{R}} (f(x-y) - f(x))k(F_h(y)/h)dyp_0(x)dx \\ &= \int_{\mathbb{R}}\int_{\mathbb{R}} \frac{(f(x - F_g^{-1}(t_2h)) - f(x))k(t_2)dtp_0(x)dx}{F'_h(F_h^{-1}(ht_2))} \\ &= \int_{\mathbb{R}} \frac{\left[\int_{\mathbb{R}} f(x - F_h^{-1}(t_2h))p_0(x)dx - \int_{\mathbb{R}} f(x)p_0(x)dx\right] k(t_2)dt_2}{F'_h(F_h^{-1}(ht_2))} \\ &= \int_{\mathbb{R}} \frac{\left[\int_{\mathbb{R}} \bar{f}(F_h^{-1}(t_2h) - x)p_0(x)dx - \int_{\mathbb{R}} \bar{f}(0 - x)p_0(x)dx\right] k(t_2)dt_2}{F'_h(F_h^{-1}(ht_2))} \\ &= \int_{\mathbb{R}} \frac{[\bar{f} * p_0(F_h^{-1}(ht_2)) - \bar{f} * p_0(0)]k(t_2)dt_2}{F'_h(F_h^{-1}(ht_2))}\end{aligned}$$

The expression on the right hand side is well defined since $f \in \mathcal{L}^p(\mathbb{R}, \lambda)$ implies that $\bar{f} \in \mathcal{L}^p(\mathbb{R}, \lambda)$ and hence $\bar{f} * p_0 \in C(\mathbb{R})$ using equation (C3).
Lemma 5 of Giné and Nickl [2008] is stated below

Lemma C.2 *(a) Let, $f \in C(\mathbb{R})$ be such that $Df$ exists and is bounded, and let $\nu \in M(\mathbb{R})$ be a signed measure. Then, for every $x \in \mathbb{R}$, $D(f * \nu)(x)$ exists and $D(f * \nu)(x) = (Df * \nu)(x)$ holds.*

*(b) Let, $g \in C(\mathbb{R})$, let $f \in \mathcal{BV}(\mathbb{R})$, and suppose that $g * f(x)$ is defined for every $x \in \mathbb{R}$. Then for every $x \in \mathbb{R}$, $D(g * f)(x)$ exists and $D(g * f)(x) = (g * \nu_f)(x)$ holds, where $\nu_f$ is the finite signed measure defined by $\nu_f((a, b]) = \tilde{f}(b) - \tilde{f}(a)$.*

**Proof of Theorem 4.1.** Since $(\mathbb{L}_n * k_{\hat{b}_n} - \mathbb{P})(f+c) = (\mathbb{L}_n * k_{\hat{h}_n} - \mathbb{P})(f)$ for any constant $c$, we may assume without loss of generality that the class $\mathcal{U}$ is uniformly bounded. Now, $\mathcal{U}$ being uniformly bounded, we have $\mathcal{U} \subseteq \mathcal{L}^1(|\mu|)$ as $k \in \mathcal{L}^1(\mathbb{R}, \lambda)$ and also $\sup_{f\in\mathcal{U}} \int_{\mathbb{R}} ||f(\cdot - y)||_{2,\mathbb{P}} d|\mu_n|(y) < \infty$. Now the following condition is proved in Part (a) Proposition 1 of Giné and Nickl [2008].

$$\sup_{f\in\mathcal{U}} E\left(\int_{\mathbb{R}} (f(X+y) - f(X))d\mu_n(y)\right)^2 \to 0 \ as \ n \to \infty. \tag{C7}$$

Now only the bias condition is required to be shown

$$\sup_{f\in\mathcal{U}} \sqrt{n}\left|E\int_{\mathbb{R}} (f(X+y) - f(X))d\mu_n(y)\right| \to 0 \ as \ n \to \infty \tag{C8}$$

If $t_1 = 0$ it follows as in the proof of Lemma C.2 (for proof see Lemma 5 of Giné and Nickl [2008]) (without limits) that $\{\bar{f} * p_0 : f \in \mathcal{U}\}$ is a bounded subset of the space of bounded Lipschitz functions on $\mathbb{R}$.
$\therefore$ Using Lemma C.1, we obtain that the bias is dominated by $C\hat{b}_n^{2-l}$ with $C$ depending only on $k$ and $\sup_{f\in\mathcal{U}} ||f||_{TV}$, which completes the proof of the theorem for $t_1 = 0$.
Next, we consider the bias for $t > 0$. By using Lemma C.2 iteratively, we have, for $f \in \mathcal{U}$,

$$D^{[t_1]+1}(p_0 * f) = D^{[t_1]}(p_0 * \nu_f) = D^{[t_1]}p_0 * \nu_f \tag{C9}$$

where $[t_1]$ denotes the integer part of $t_1$. In particular, $D^{\alpha}(p_0 * f)$ exists and is a bounded and continuous function for every $0 \le \alpha \le [t_1] + 1$. We consider two cases:
(a) When $t_1 = [t_1]$ is an integer, we have from (C9) and (C4) that

$$\sup_{f\in\mathcal{U}} ||D^{t_1+1}(p_0 * f)||_\infty \le ||D^{t_1}p_0||_\infty \sup_{f\in\mathcal{U}} ||\nu_f|| < \infty, \tag{C10}$$

where $||\nu_f|| \le ||f||_{TV}$ (see equation (C6)).
(b) When $t_1 > [t_1]$ is non integer we show that $D^{[t_1]+1}(p_0 * f)$ is Hölder-continuous of order $t_1 - [t_1]$. We have

$$\begin{aligned} |h|^{[t_1]-t_1} \ & |D^{[t_1]+1}(p_0 * f)(x+h) - D^{[t_1]+1}(p_0 * f)(x)| \\ = \quad & \left|\int_{\mathbb{R}} |h|^{[t_1]-t_1}(D^{[t_1]}p_0(x+y+h) - D^{[t_1]}p_0(x+y))d\nu_f(y)\right| \\ \le \quad & ||D^{[t_1]}p_0||_{t_1-[t_1],\infty}||f||, \end{aligned} \tag{C11}$$

which is bounded uniformly in $f \in \mathcal{U}$, since $p_0 \in C^t(\mathbb{R})$ implies $D^{[t_1]}p_0 \in C^{t_1-[t_1]}(\mathbb{R})$ and since the variation of $\nu_f$ is uniformly bounded (using $||\nu_f|| \le ||f||_{TV}$).

We use Lemma C.1 to bound the bias term. If $f \in \mathcal{U}$ then also $\bar{f} \in \mathcal{U}$. We consider first $2t_1 + 1 - l$ noninteger. Then by Taylor expansion and change of variables we have

$$
\begin{aligned}
|E(f * k(X) - f(X))| = \Big| \int_{\mathbb{R}} \frac{k(s)}{(F'_{\hat{b}_n}(F^{-1}_{\hat{b}_n}(\hat{b}_n s)))} \Big[ & D(p_0 * \bar{f})(0) s \hat{b}_n + \cdots \\
& + \frac{1}{[2t_1 + 1 - l]!} D^{[2t_1+1-l]}(p_0 * \bar{f})(0) s^{[2t_1+1-l]} \hat{b}_n^{[2t_1+1-l]} \Big] ds \\
& + \frac{\hat{b}_n^{[2t_1+1-l]}}{[2t_1 + 1 - l]! F'_{\hat{b}_n}(F^{-1}_{\hat{b}_n}(\hat{b}_n s))} \int_{\mathbb{R}} k(s) s^{[2t_1+1-l]} \Big[ D^{[2t_1+1-l]}(p_0 * \bar{f})(F^{-1}_{\hat{b}_n}(\zeta \hat{b}_n s)) \\
& - D^{[2t_1+1-l]}(p_0 * \bar{f})(0) \Big] ds \Big| \; where \; F'_{\hat{b}_n}(F^{-1}_{\hat{b}_n}(\hat{b}_n s)) \neq 0 \qquad \text{(C12)}
\end{aligned}
$$

where $0 < \zeta < 1$. The first $[2t_1 + 1 - l]$ terms in the above display are all equal to zero by the choice of the kernel (see Definition 3). From (C11) and (C10) we have $D^{[2t_1+1-l]}(p_0 * \bar{f})$ is contained in $C^{\alpha}(\mathbb{R})$ where $\alpha = 2t_1 + 1 - l - [2t_1 + 1 - l] > 0$, and again using the assumptions on the kernel, we have that the last term in (C12) is bounded in absolute value by

$$
C \hat{b}_n^{2t_1+2-l} \int_{\mathbb{R}} \frac{|k(s)||s|^{2t_1+1-l}}{\left| \sum_{i=1}^{n} k\left( \frac{F^{-1}_{\hat{b}_n}(s\hat{b}_n) - X_i}{\hat{b}_n} \right) \right|},
$$

the constant $C$ being equal to

$$
C = \frac{n}{[2t_1 + 1 - l]!} \sup_{f \in \mathcal{U}} ||D^{[t_1+1-l]}(p_0 * f)||_{\infty,\alpha} < \infty.
$$

Now, by the choice of $\hat{b}_n$, we obtain the bound

$$
\sup_{f \in \mathcal{U}} \sqrt{n} |E(f * k(X) - f(X))| = O(\sqrt{n} \hat{b}_n^{2t_1+2-l}) = o(1) \qquad \text{(C13)}
$$

for the bias term. When $2t_1 + 1 - l$ is integer,

$$
\begin{aligned}
|E(f * k(X) - f(X))| = \Big| \int_{\mathbb{R}} \frac{k(s)}{(F'_{\hat{b}_n}(F^{-1}_{\hat{b}_n}(\hat{b}_n s)))} [ & D(p_0 * \bar{f})(0) s \hat{b}_n + \cdots \\
& + \frac{1}{(2t_1 + 1 - l)!} D^{2t+1-l}(p_0 * \bar{f})(0) s^{2t_1+1-l} \hat{b}_n^{2t_1+1-l} ] ds \\
& + \frac{\hat{b}_n^{2t_1+1-l}}{(2t_1 + 1 - l)! (F'_{\hat{b}_n}(F^{-1}_{\hat{b}_n}(\hat{b}_n s)))} \int_{\mathbb{R}} k(s) s^{2t_1+1-l} \\
& [ \; D^{2t_1+1-l}(p_0 * \bar{f})(F^{-1}_{\hat{b}_n}(\zeta \hat{b}_n s)) ] ds \Big| \qquad \text{(C14)}
\end{aligned}
$$

for some $0 < \zeta < 1$. By using the choice of the kernel and since $D^{2t_1+1-l}(p_0 * \bar{f})$ is sup-norm bounded uniformly in $f \in \mathcal{U}$ (from (C11) and (C10)) and hence we get the equation (C13).

**Proof of Theorem 4.3.** Now, for $0 \leq t_1 \leq 1$, we define $\psi_n(t_1) = -\int_{t_1}^{1} J_n dF$ so that $\frac{c_{ni}}{n} = \left[ \psi_n\left(\frac{i}{n}\right) - \psi_n\left(\frac{(i-1)}{n}\right) \right]$ for $0 \leq t_1 \leq 1$.

$$W_n - \mu_n = \int_0^1 g(\widetilde{F}_n^{-1}(t_1))d\psi_n(t_1) - \int_0^1 g(t_1)d\psi_n(t_1)$$
$$= \int_0^1 g(t_1)d[\psi_n(\widetilde{F}_n(t_1)) - \psi_n(t_1)]$$

Integrating by parts, we have

$$W_n - \mu_n = \lim_{\theta \to 0}[g(t_1)\{\psi_n(\widetilde{F}_n(t_1)) - \psi_n(t_1)\}]_{\theta}^{1-\theta} - \int_0^1 [\psi_n(\widetilde{F}_n(t_1)) - \psi_n(t_1)]dg(t_1) \tag{C15}$$

Now, for $0 < t_1 < \xi_{n1}$, we have

$$|g(t_1)\{\psi_n(\widetilde{F}_n(t_1)) - \psi_n(t_1)\}| \leq D(t_1)\int_0^{t_1} B(u)du \leq M{t_1}^{\delta} \to 0 \ as \ t_1 \to 0$$

A similar argument for $\xi_{nn} < t_1 < 1$ holds. Thus equation (C15) can be written as

$$W_n - \mu_n = -\int_0^1 [\psi_n(\widetilde{F}_n(t_1)) - \psi_n(t_1)]dg(t_1)$$

Now, we can write the above equation as

$$W_n - \mu_n = -\gamma_n - S_n, \tag{C16}$$

where $\gamma_n = \int_0^1 [\psi_n(\widetilde{F}_n(t_1)) - \psi_n(t_1) - \{\widetilde{F}_n(t_1) - t_1\}J(t_1)]dg(t_1)$ and $S_n = \int_0^1 [\widetilde{F}_n(t_1) - t_1]J(t_1)dg(t_1)$.
Now,

$$\sqrt{n}\gamma_n = \int_0^1 U_n(t_1)A_n(t_1)dg(t_1), \tag{C17}$$

where $U_n(t_1) = \sqrt{n}(\widetilde{F}_n(t_1) - t_1)$ and $A_n(t_1) = \frac{1}{\widetilde{F}_n(t_1)-t_1}\int_{t_1}^{\widetilde{F}_n(t_1)} J_n(u)du - J(t_1)$.
Now,

$$\sqrt{n}|\gamma_n| \leq ||U_n/h|| \int_0^1 |A_n(t_1)|h(t_1)d|g|(t_1),$$

where $||\cdot||$ denotes the sup-norm on $(0,1)$. Now, for $\xi_{n1} \leq t_1 \leq \xi_{nn}$, it follows from Assumption(A) that

$$|A_n(t_1)| \leq \frac{1}{\widetilde{F}_n(t_1) - t_1}\int_{t_1}^{\widetilde{F}_n(t_1)} |J_n(u)|du + |J(t_1)|$$
$$\leq B(t_1) \vee B(\widetilde{F}_n(t_1)) + B(t_1)$$

From Theorem B.5, $\exists$ a set $A \subset \Omega$ with $P(A) = 1$ for a significantly large $n$. So for $\xi_{n1} \leq t_1 < \xi_{nn}$ we have

$$|A_n(t_1)|h(t_1) \leq M{t_1}^{1-\delta/2-b_1(1+\tau_1)}(1-t_1)^{1/2-\delta/2-b_2(1+\tau_2)}.$$

Now for $0 < t_1 < \xi_{n1}$ and $\widetilde{F}_n(t_1) = 0$, we have

$$|A_n(t_1)| \leq \frac{1}{t_1}\int_0^1 B(u)du + B(t_1) \leq M{t_1}^{-b_1}.$$

Similarly, we have for $\xi_{nn} \leq t_1 < 1$.

$\therefore$ On a set $A$ we have

$$|A_n(t_1)|h(t_1) \leq M{t_1}^{1-\delta/2-b_1(1+\tau_1)}(1-t)^{1/2-\delta/2-b_2(1+\tau_2)}.$$

Now, from Assumption(A) we see that the right-hand side of equation (C17) is $|g|$-integrable and by Assumption(B) we have $A_n(t) \to 0$, $|g|$-a.e as $n \to \infty$ with probability one.

Hence we can write $\sqrt{n}|\gamma_n| \to 0$ as $n \to \infty$ with probability one.

$\therefore$ Equation (C16) can be written as

$$\sqrt{n}(W_n - \mu_n) = -\sqrt{n}S_n. \tag{C18}$$

Now,

$$\sqrt{n}S_n = \int_0^1 \mathbb{G}_n(t_1)J(t_1)dg(t_1),$$

where $\mathbb{G}_n(t_1) = \sqrt{n}(\widetilde{F}_n(t_1) - t_1)$. We now define

$$S = \int_0^1 \mathbb{G}(t_1)J(t_1)dg(t_1)$$

so that $S$ is a $N(0, \sigma^2)$ random variable and $\sigma^2$ is finite by Assumption(A).

Now using Corollary 2 and dominated convergence theorem, we can write that

$$\sqrt{n}S_n = \int_0^1 \mathbb{G}_n(t_1)J(t_1)dg(t_1) \xrightarrow{\text{d}} S = \int_0^1 \mathbb{G}(t_1)J(t_1)dg(t_1)$$

And hence we can write

$$\sqrt{n}(W_n - \mu_n) \xrightarrow{\text{d}} -\int_0^1 \mathbb{G}(t_1)J(t_1)dg(t_1).$$

**Proof of Theorem 4.5.** The proof is similar to the proof of Theorem 2.3 of Degenhardt et al. [1996].

**Proof of Theorem 4.6.** The proof is similar to the proof of Theorem 4.3 and using Theorem 4.5 we prove the result.

**Proof of Proposition 7.1.** To estimate the expectation of $X^N_{SR}(\phi)$, we first consider a single trading day failure $X^{(i)}_{SR}(\phi)$ and estimate the expectation of $X^{(i)}_{SR}(\phi)$.

$$
\begin{aligned}
E[X_{SR}^{(i)}(\phi)] &= E\Big[\int_0^1 \phi(p)K\Big(\frac{F_{n,b}(VaR_i^p - F_{n,b}(L_i))}{b}\Big)dp\Big] \\
&= \int_0^1 \phi(p)E\Big[K\Big(\frac{F_{n,b}(VaR_i^p - F_{n,b}(L_i))}{b}\Big)\Big]dp \\
&= \int_0^1 \phi(p)\Big[F(VaR_i^p) + \frac{b^4\mu_2^2(k)}{4}\Big(\frac{f'(VaR_i^p)f''(VaR_i^p)}{f^3(VaR_i^p)} - \frac{f'''(VaR_i^p)}{f^2(VaR_i^p)}\Big)\Big]dp \\
&= \int_0^1 \phi(p)pdp
\end{aligned}
$$

Then,

$$
\begin{aligned}
E[X_{SR}^N(\phi)] &:= E\Big[\frac{1}{N}\sum_{i=1} X_{SR}^{(i)}(\phi)^N\Big] \\
&= \frac{1}{N}\sum_{i=1}^{N} E[X_{SR}^{(i)}(\phi)] \\
&= \int_0^1 \phi(p)pdp
\end{aligned}
$$

Hence the expected value of the average over N trading days is equal to the expected value of a single trading day. Again to estimate the variance of $X_{SR}^N(\phi)$, we consider the single trading day failure $X_{SR}^{(i)}(\phi)$ and estimate the variance of $X_{SR}^{(i)}(\phi)$.

$$
\begin{aligned}
E[(X_{SR}^{(i)}(\phi))^2] &= E\Big[\Big(\int_0^1 \phi(p)K\Big(\frac{F_{n,b}(VaR_i^p) - F_{n,b}(L_i)}{b}\Big)dp\Big)^2\Big] \\
&= E\Big[\Big(\int_0^1 \phi(p)K\Big(\frac{p - F_{n,b}(L_i))}{b}\Big)dp\Big)^2\Big] \\
&= E\Big[\int_0^1 \phi(p)K\Big(\frac{p - F_{n,b}(L_i))}{b}\Big)dp\int_0^1 \phi(q)K\Big(\frac{q - F_{n,b}(L_i))}{b}\Big)dq\Big] \\
&= 2E\Big[\int_0^1\int_0^p \phi(p)\phi(q)K\Big(\frac{p - F_{n,b}(L_i))}{b}\Big)K\Big(\frac{q - F_{n,b}(L_i))}{b}\Big)dpdq\Big] \\
&= 2E\Big[\int_0^1\int_0^p \phi(p)\phi(q)K\Big(\frac{q - F_{n,b}(L_i))}{b}\Big)dpdq\Big],\ q \le p \\
&= 2\int_0^1\int_0^p \phi(p)\phi(q)qdpdq
\end{aligned}
$$

Now,

$$
\begin{aligned}
V[X_{SR}^{(i)}(\phi)] &:= E[(X_{SR}^{(i)}(\phi))^2] - E[X_{SR}^{(i)}(\phi)]^2 \\
&= 2\int_0^1\int_0^p \phi(p)\phi(q)qdpdq + \Big(\int_0^1 \phi(p)pdp\Big)^2
\end{aligned}
$$

Therefore,

$$
\begin{aligned}
V[X_{SR}^N(\phi)] &= V\Big[\frac{1}{N}\sum_{i=1}^{N} X_{SR}^{(i)}(\phi)\Big] \\
&= \frac{1}{N^2}\Big(\sum_{i=1}^{N} V[X_{SR}^{(i)}(\phi)] + \sum_{i\neq j} \underbrace{corr\langle X_{SR}^{(i)}(\phi),\ X_{SR}^{(j)}(\phi)\rangle}_{=0\ under\ H_0}\Big) \\
&= \frac{1}{N^2}\sum_{i=1}^{N}\Big[2\int_0^1\int_0^p \phi(p)\phi(q)qdpdq - \Big(\int_0^1 \phi(p)pdp\Big)^2\Big] \\
&= \frac{1}{N}\Big(2\int_0^1\int_0^p \phi(p)\phi(q)qdpdq - \Big(\int_0^1 \phi(p)pdp\Big)^2\Big)
\end{aligned}
$$

**Proof of Lemma 7.2.** Under the null hypothesis, the sequence $\{X_{SR}^{(i)}\}_{i=1}^N$ is i.i.d with bounded mean $\mu_\phi$ and variance $\sigma_\phi^2$ given by equation (15) and (16) respectively. Thus by the Lindeberg-Levy Central Limit Theorem we have

$$\sqrt{N}(X_{SR}^N(\phi) - \mu_\phi) \xrightarrow[N\to\infty]{\mathrm{d}} \mathcal{N}(0, \sigma_\phi^2). \tag{C19}$$

$\therefore$ for large $N$, $X_{SR}^N$ is approximately normal and hence admits a Z-test.

**Proof of Theorem 7.3.** By Lemma 7.2 the test statistic $X_{SR}^N$ admits a Z-test with standard Z-score. By Proposition 7.1 the mean and variance of $X_{SR}^N$ under the null hypothesis are given by equation (15) and (16) and hence we obtain equation (17).